\documentclass[useAMS,usenatbib]{mn2e}
\usepackage[pdftex]{graphicx}

\usepackage{astro_journal_names}
\defcitealias{2004MNRAS.347..255B}{BT04} 

\title[Are peculiar velocity surveys competitive as a cosmological probe?]{Are peculiar velocity surveys competitive as a cosmological probe?} 
\author[J. Koda et al.]{Jun Koda,$^{1,2}$\thanks{E-mail: jkoda@astro.swin.edu.au}
  Chris Blake,$^{1,2}$
  Tamara Davis,$^{2,3,4}$
  Christina Magoulas,$^{5}$\newauthor 
  Christopher M. Springob,$^{2,6,7}$
  Morag Scrimgeour,$^{2,6}$
  Andrew Johnson,$^{1,2}$\newauthor
  Gregory B. Poole,$^{5}$
  Lister Staveley-Smith$^{2,6}$\\
  $^1$ Centre for Astrophysics \& Supercomputing, Swinburne University of 
       Technology, PO Box 218, Hawthorn, VIC 3122, Australia  \\
  $^2$ ARC Centre of Excellence for All-sky Astrophysics (CAASTRO)  \\
  $^3$ School of Mathematics and Physics, University of Queensland, Brisbane, QLD 4072, 
       Australia\\
  $^4$ Dark Cosmology Centre, Niels Bohr Institute, University of Copenhagen, 
       Juliane Maries Vej 30, DK-2100 Copenhagen, Denmark\\
  $^5$ School of Physics, University of Melbourne, Parkville, VIC 3010 Australia\\
  $^6$ International Centre for Radio Astronomy Research, 
       University of Western Australia, 35 Stirling Highway, 
       Perth WA 6009, Australia\\
  $^7$ Australia Astronomical Observatory, P.O. Box 915, North Ryde, NSW 1670, Australia
}

\usepackage{color}

\begin{document}
\label{firstpage}

\maketitle

\begin{abstract}
  Peculiar velocity surveys, which measure galaxy peculiar velocities
  directly from standard candles in addition to redshifts, can provide
  strong constraints on the linear growth rate of cosmological
  large-scale structure at low redshift.  The improvement originates
  from the physical relationship between galaxy density and peculiar
  velocity, which substantially reduces cosmic variance. We present
  the results of Fisher matrix forecasts of correlated fields of
  galaxy density and velocity. Peculiar velocity can improve the
  growth rate constraints by about a factor of two compared to density
  alone for surveys with galaxy number density of about $10^{-2}
  (h^{-1} \mathrm{Mpc})^{-3}$, if we can use all the information for
  wavenumber $k \le 0.2\,h\mathrm{Mpc}^{-1}$. In the absence of accurate
  theoretical models at $k = 0.2\,h \textrm{Mpc}^{-1}$, the
  improvement over redshift-only surveys is even larger --- around a
  factor of 5 for $k \le 0.1\,h\mathrm{Mpc}^{-1}$. Future peculiar
  velocity surveys, TAIPAN, and the all-sky H\,{\sc i} surveys, WALLABY
  and WNSHS, can measure the growth rate, $f\sigma_8$ to 3 per cent at
  $z\sim0.025$. Although the velocity subsample is about an order of
  magnitude smaller than the redshift sample from the same survey, it
  improves the constraint by 40 per cent compared to the same survey
  without velocity measurements. Peculiar velocity surveys can also
  measure the growth rate as a function of wavenumber with 15--30 per
  cent uncertainties in bins with widths $\Delta k = 0.01
  \,h\mathrm{Mpc}^{-1}$ in the range $0.02 \,h \mathrm{Mpc}^{-1} \la k
  \le 0.1 \,h \mathrm{Mpc}^{-1}$, which is a large improvement over
  galaxy density only. Such measurements on very large scales can
  detect signatures of modified gravity or non-Gaussianity through
  scale-dependent growth rate or galaxy bias.  We use $N$-body
  simulations to improve the modelling of auto- and cross-power
  spectra of galaxy density and peculiar velocity by introducing a new
  redshift-space distortion term to the velocity, which has been
  neglected in previous studies. The velocity power spectrum has a
  damping in redshift space, which is larger than that naively
  expected from the similar effect in the galaxy power spectrum.
\end{abstract}


%
% 1. Introduction
%
\section{Introduction}
\label{sec:introduction}
Whether the accelerated expansion of the Universe and the growth of
the large-scale structure can be fully explained by the standard
Lambda-Cold Dark Matter ($\Lambda$CDM) model, especially by the
cosmological constant $\Lambda$, is one of the main questions of modern
cosmology. Dark energy, which can have an equation of state
different from the cosmological constant, or theories of gravity alternative to
General Relativity, might be a source of recent accelerated
expansion. Galaxy peculiar velocities provide powerful tests of the
$\Lambda$CDM model through measurements of the linear growth rate,
which are complementary to other cosmic probes. The linear growth
rate, $f \equiv d\ln D(a)/d\ln a$, is the logarithmic derivative of
the linear growth factor $D$ with respect to the cosmic scale factor
$a$. Different models of dark energy or modified gravity give
different growth rates as a function of time or scale.

Observed redshift is a combined effect of cosmological expansion which
depends on the distance to the source, and the Doppler shift which
depends on the peculiar velocity of the light source. Direct
measurements of line-of-sight peculiar velocities from redshifts and
distances determined by the Tully-Fisher relation, Fundamental Plane,
or supernovae have a long history \citep[see][for a
  review]{1995PhR...261..271S}. Latest peculiar velocity surveys
have increased their samples to about 5000 Tully-Fisher velocities
\citep{2007ApJS..172..599S}, and 9000 Fundamental-Plane velocities
\citep{2012MNRAS.427..245M}. Supernovae samples are also competitive
because the smaller sample size is compensated by their better
precision per measurement \citep{2012MNRAS.420..447T,
  2013arXiv1310.4184F}. Future surveys plan to expand peculiar
velocity samples further, as we present in this paper. All of these
samples are limited to low redshift ($z \la 0.05$), because the
velocity error grows at least linearly with distance, 8--25 per cent of
the Hubble recession velocity, due to intrinsic scatter in distance estimation.

A completely different path to measure peculiar velocity is the
kinetic Sunyaev-Zel'dovich effect \citep{1980MNRAS.190..413S}, which
measures line-of-sight velocities of galaxies or clusters of galaxies
with respect to the cosmic microwave background (CMB). The bulk motion
of electrons along the large-scale peculiar velocity field make a tiny
contribution to the CMB temperature via Thomson scattering between
CMB photons and free electrons in galaxies or clusters. The kinetic
Sunyaev-Zel'dovich effect has just started to be measurable
\citep{2012PhRvL.109d1101H, 2013MNRAS.430.1617L}, and has the potential
to provide peculiar velocity measurements  with errors that do not diverge
linearly with distance.

Alternatively, the information of line-of-sight peculiar velocity is
also encoded in the anisotropic pattern of large-scale galaxy
clustering, known as the redshift-space distortion
\citep{1987MNRAS.227....1K}. Many measurements of growth rates through
redshift-space distortion have been made across wide ranges of
redshifts \citep[e.g.,][]{2001Natur.410..169P, 2004PhRvD..69j3501T,
  2008Natur.451..541G, 2011MNRAS.415.2876B, 2012MNRAS.423.3430B,
  2013MNRAS.429.1514S, 2013A&A...557A..54D}. These measurement are, so
far, all consistent with the $\Lambda$CDM Universe, and continue to
improve in precision and redshift range
\citep[e.g.,][]{2009MNRAS.397.1348W, 2013PhR...530...87W,
  2013LRR....16....6A, 2013arXiv1308.6070D}. In this paper, we refer to
these measurements as those from `galaxy density only',
`redshift-space distortions only', or `redshift only',
interchangeably, to distinguish them from peculiar velocity
surveys.

CMB measurements constrain cosmological parameters precisely within the
$\Lambda$CDM cosmological model, but are not as sensitive to
low-redshift growth of structure. Peculiar velocity surveys therefore
allow a powerful consistency check of scenarios of dark energy or
modified theories of gravity. 
Low-redshift data are useful as these correspond to the epoch when the
ratio of dark energy density to Critical density is the largest and
where any deviation from $\Lambda$CDM is likely to be the most
significant \citep{2012ApJ...751L..30H}.

The amplitude of velocity fluctuation measures $f\sigma_8$ where
$\sigma_8$ is the amplitude of matter perturbation smoothed on spheres
of $8h^{-1}\mathrm{Mpc}$, and $h$ is the Hubble constant in units of
$100 \textrm{ km s}^{-1}\mathrm{Mpc}^{-1}$. The velocity power spectrum
\citep{1995ApJ...455...26J, 2009MNRAS.400.1541A, 2011MNRAS.414..621M}
measures $f\sigma_8$ as a function of wavenumber, and the bulk flow
\citep{1988MNRAS.231..149K, 2009MNRAS.392..743W} or low-order moments
\citep{2010MNRAS.407.2328F} of the velocity field are a cross-check of the 
model
at large scales. Many bulk flow measurements show that the local
velocity fluctuations at large scales are larger than found at a typical
location in the $\Lambda$CDM Universe, but whether the large bulk flow
is inconsistent with the $\Lambda$CDM Universe
\citep{2009MNRAS.392..743W, 2010MNRAS.407.2328F} or consistent
\citep{2011ApJ...736...93N, 2012MNRAS.420..447T, 2013MNRAS.428.2017M}
is still under debate. Since we can
only measure the large-scale bulk flow around us, it is not decisive
(from velocity data alone) whether the large-scale velocities are a
problem of the $\Lambda$CDM universe or a statistical outcome that we
are simply in a high velocity region of the $\Lambda$CDM Universe
(cosmic variance).

One of the advantages of peculiar velocity surveys is that they have
multiple tracers, galaxy density and peculiar velocity, which can be
used to measure growth rate beyond the cosmic variance limit. In
  contrast, methods that use only one tracer of large-scale
structure are limited by the number of fluctuation modes in the
observed volume. In the case of peculiar velocity surveys, cosmic variance
can be reduced by first predicting the expected velocity field from
the galaxy distribution \citep{1994ApJ...421L...1N,
  2002MNRAS.335...53B, 2006MNRAS.373...45E, 2008MNRAS.389..497K,
  2010ApJ...709..483L, 2012MNRAS.425.2422K}, and then comparing
the model velocities with the observed velocities. The ratio of the two
gives the $\beta \equiv f/b$ parameter, where $b$ is the galaxy bias
\citep{1996ApJ...473...22D, 2001MNRAS.326.1191B, 2011MNRAS.413.2906D,
  2012MNRAS.425.2880M}. Because the reconstructed velocity and the
observed velocity share the same random perturbation, the measurement
of $\beta$ is not limited by the cosmic variance, but continues to
improve as the statistical error of velocities is reduced by
increasing the number of velocity measurements. To obtain a more
fundamental quantity, which does not depend on galaxy selection
through bias $b$, one can obtain $f\sigma_8$ by multiplying $\beta$ by
the amplitude of the galaxy clustering, $b\sigma_8$. Similar
techniques for going beyond the cosmic variance limit have also been
proposed by using multiple populations of galaxy densities with
different biases instead of density and velocity
\citep{2009JCAP...10..007M, 2010MNRAS.407..772G, 2011MNRAS.416.3009B}.

The dependence of the growth rate on the wavenumber $k$ is also an
important observable for distinguishing theories of gravity
\citep[e.g., ][]{2009JCAP...10..004S, 2012MNRAS.425.2128J,
  2013PhRvD..87b3501A, 2013JCAP...08..029A,
  2013arXiv1309.6783T}. Although the overall normalisation of $\beta$
depends on galaxy bias $b$, we can test whether $\beta$ is independent
of $k$, as predicted by General Relativity. Similarly, scale
dependence of $\beta$ can also constrain primordial non-Gaussianity
which generates scale-dependent bias \citep{2008PhRvD..77l3514D,
  2008ApJ...677L..77M, 2008JCAP...08..031S, 2009MNRAS.396...85D,
  2012PhRvD..86f3526A, 2013MNRAS.428.2765D}. The advantage of multiple
tracers over a single tracer become larger for testing the scale
dependence than for the case of assuming constant $\beta$, improving
the constraints especially at low $k$ by evading the cosmic variance
\citep{2009PhRvL.102b1302S, 2011PhRvD..84h3509H, 2013MNRAS.tmp.2399M}.

The purpose of this paper is to investigate whether future
peculiar velocity surveys are competitive with redshift surveys, and
evaluate how much the peculiar velocity data improve the constraints
on the growth rates $f\sigma_8$ and
$\beta$. \citet{2004MNRAS.347..255B} introduced the Fisher matrix
analysis to peculiar velocity surveys, and forecast the performance
of the 6dF Galaxy Survey Peculiar Velocity Survey
\citep[6dFGSv,][]{2009MNRAS.399..683J, 2012MNRAS.427..245M}. Since
their paper simply uses the linear velocity power spectrum without
redshift-space distortions, we first improve the model equations for
auto- and cross-power spectra of galaxy density and peculiar velocity
by comparing the equations with an $N$-body simulation in
Section~\ref{sec:velocity_powerspectrum}. Using the model, we show how
constraints from peculiar velocity surveys improve compared with those
from redshift surveys, using the Fisher matrix formalism in a general
case in Section~\ref{sec:fisher}. We present our new forecast for
future peculiar velocity surveys in Section~\ref{sec:forecast}, and
summarise the results in Section~\ref{sec:summary}. Throughout the
paper, we use a flat $\Lambda$CDM cosmology with $\Omega_m=0.273$,
$\Omega_\Lambda = 0.727$, $\Omega_b=0.0546$, $h=0.705$,
$\sigma_8=0.812$, and $n_s= 0.961$.

%
% 2. Velocity Power Spectrum in redshift space
%
\section{Velocity Power Spectrum in Redshift Space}
\label{sec:velocity_powerspectrum}
In this section, we introduce simple model equations that describe
the auto- and cross-power spectra of galaxy number density and
line-of-sight velocity \textit{in redshift space}. Although real-space
distances to galaxies are in principle measurable in peculiar velocity
surveys, we assume that analyses are performed in redshift space. This is
because the velocity measurement errors are sufficiently large that
the real-space location has a significant error, complicating
clustering measurements, whereas the redshift-space position is
accurately known.  We use these model equations to calculate the Fisher
matrix in Section~\ref{sec:fisher}, assuming the density and
velocities are Gaussian random fields completely characterised by the
power spectra.

We denote the galaxy density contrast field by $\delta_g$, the
velocity vector field by $\bmath{v}$, and the line-of-sight velocity
by $u$, respectively. We focus on the line-of-sight velocity, instead
of the velocity vector or the velocity divergence, because the
line-of-sight velocity is the observable in peculiar velocity
surveys. Throughout the paper, we assume the flat-sky approximation,
such that the line of sight is fixed to the third axis: $u \equiv
v_3$. The auto- and cross-power spectra of $\delta$ and $u$ are
ensemble averages of their products in Fourier space:
  $P_{gg}(\bmath{k}) =
       V^{-1} \langle \delta_g(\bmath{k}) \delta_g(\bmath{k})^* \rangle$,
  $P_{gu}(\bmath{k}) = 
       V^{-1} \langle \delta_g(\bmath{k}) u(\bmath{k})^* \rangle$,
and
  $P_{uu}(\bmath{k}) =
       V^{-1} \langle u(\bmath{k}) u(\bmath{k})^* \rangle$,
where $V$ is a volume of a periodic box, and $\delta_g(\bmath{k})$ and
$u(\bmath{k})$ are the Fourier transform of $\delta_g(\bmath{x})$
and $u(\bmath{x})$, respectively, for a convention that the Fourier
transformation of a function $f(\bmath{x})$ is
  $f(\bmath{k}) = \int_V f(\bmath{x}) e^{-\mathrm{i}\bmath{k}\cdot\bmath{x}} d^3x$.

Although the cross-power spectrum is generally a complex-valued function,
the cross-power of $\delta_g$ and $u$ has a purely imaginary value by parity
invariance. If you flip the Universe to the mirror image, $\bmath{x}
\mapsto -\bmath{x}$ and $u \mapsto -u$, the ensemble averaged quantities,
including the cross power, must be the same, because the statistical
property of the initial condition and the time-evolution by gravity are
both indistinguishable under this parity transformation. Fourier modes
transform as
  $\delta_g(\bmath{k})\mapsto \delta_g(-\bmath{k})$
and
  $u(\bmath{k}) \mapsto -u(-\bmath{k})$,
respectively, under the parity transformation. This argument leads to,
$V P_{gu}(\bmath{k}) 
= \langle \delta_g(\bmath{k}) u(\bmath{k})^* \rangle 
= -\langle \delta_g(-\bmath{k}) u(-\bmath{k})^*\rangle 
= -\langle \delta_g^*(\bmath{k}) u(\bmath{k}) \rangle 
= -V P_{gu}(\bmath{k})^*$,
where the reality condition $f(-\bmath{k}) = f(\bmath{k})^*$, for any
real function $f(\bmath{x})$, is used. This shows that $P_{gu}$ is
pure imaginary. Note that this is true for the ensemble average with
infinite volume or infinite number of random realisations; $P_{gu}$
estimated from a finite number of modes is consistent with pure
imaginary only within statistical uncertainty. This property of having
either a real or a pure imaginary off-diagonal element is not limited
to linear perturbation theory, but also applicable to non-linear power
spectra.

Since the vorticity $\bmath{\nabla}\times\bmath{v}$ is negligible on
large scales \citep{2009PhRvD..80d3504P}, the line-of-sight velocity
field is directly related to the velocity divergence field,
$\theta(\bmath{x}) \equiv -\bmath{\nabla} \cdot
\bmath{v}(\bmath{x})/(a H f)$, to a good approximation on the scales we
are interested in, where $H=H(z)$ is the Hubble parameter at redshift
$z$. The corresponding relation in Fourier space is,
\begin{equation}
  \label{eq:u_theta}
  u(\bmath{k}) = -\mathrm{i} aHf\mu \theta(\bmath{k})/k,
\end{equation}
where $k \equiv |\mathbf{k}|$, and $\mu$ is the cosine of the angle
between $\bmath{k}$ and the line of sight, $\mu 
\equiv k_3/k$. The continuity equation relates the velocity divergence
field to the time derivative of the density field. The $\theta$
variable is defined such that it is equal to $\delta$ in the linear
limit: $\theta(\bmath{k}) = \delta_m(\bmath{k})$. We review the
$N$-body simulation in Section~\ref{sec:vpower_simulation}, and
present the power spectra in real space in
Section~\ref{sec:realspace-power}. In Section~\ref{sec:zspace-power},
we introduce the model equations in redshift space, and test those
equations using the simulation.

\subsection{The Simulation}
\label{sec:vpower_simulation}
We use subhaloes in the GiggleZ simulation (Poole et al, in
preparation) to calculate simulated galaxy and velocity power
spectra. The simulation has $2160^3$ $N$-body particles of masses
$7.5\times10^9 h^{-1} M_\odot$, in a periodic box of $1 h^{-1}\textrm{
  Gpc}$ on a side. The cosmological parameters used in this simulation
are listed at the end of Section~\ref{sec:introduction}.  The
simulation is performed with the \textsc{GADGET2} code
\citep{2005MNRAS.364.1105S}, and haloes and subhaloes are found by the
\textsc{SUBFIND} code \citep{2001MNRAS.328..726S}.  We mainly present
the results for subhaloes in the mass range $10^{11.5}-10^{12} h^{-1}
M_\odot$, which roughly correspond to disk galaxies observed in
H\,{\sc i} surveys, but our results are qualitatively the same for
other subhaloes.

We compute the subhalo density field on a grid with the Cloud-in-Cell
(CIC) method. The computation of the velocity field has technical
difficulties that do not exist for the density field; for example,
using the CIC method for the velocity would cause a problem of
undefined velocities when the density of the cell is zero \citep[see,
  e.g.,][]{2009PhRvD..80d3504P, 2012MNRAS.427L..25J}. We assign the
line-of-sight velocity field on a $512^3$ grid using the
nearest particle method; i.e., for each grid point, we assign the
velocity of the nearest particle to the grid.  We Fast Fourier
Transform the grids and calculate their auto- and cross-power
spectra. We subtract shot noise from the galaxy auto-power spectrum, and
correct for the smoothing and aliasing caused by gridding. We explain
the details of this process in
Appendix~\ref{sec:appendix_power_spectra}, which enables us to
construct reliable power spectra and hence the model equations.
Numerical errors caused by gridding are less than one per cent for
$k\le 0.2 \,h\mathrm{Mpc}^{-1}$ after the corrections; our density and
velocity power spectra are sufficiently accurate for our purpose.  Our
code for calculating the velocity power spectrum with the nearest particle
method is publicly available at
\texttt{https://github.com/junkoda/np\_vpower}.

\begin{figure*}
\centering
\includegraphics[width=174mm]{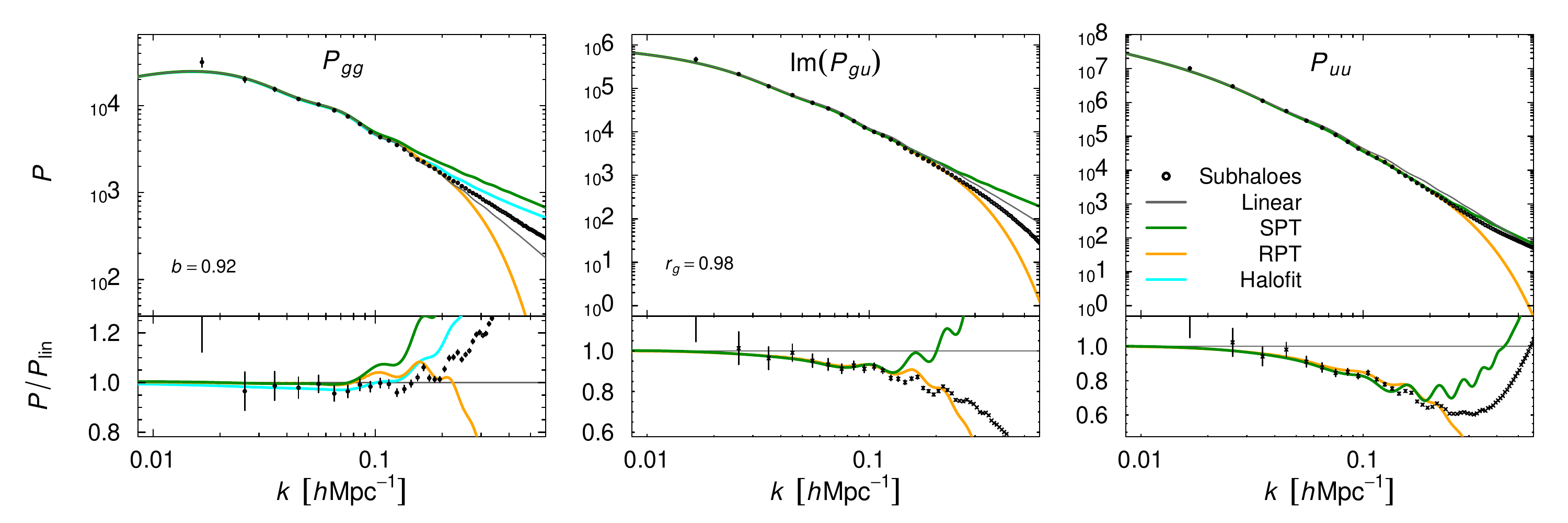}
\caption{The angle-averaged auto- and cross-power spectra for galaxy
  density and line-of-sight peculiar velocity in real space. The three
  panels for $P_{gg}$, $P_{gu}$, and $P_{uu}$ are for the galaxy
  auto-power spectrum, galaxy-velocity cross-power spectrum, and
  velocity auto-power spectrum, respectively. The points are from
  subhaloes in the GiggleZ simulation, grey lines are linear
  theory, green lines are one-loop standard perturbation theory (SPT),
  and orange lines are 1-loop renormalised perturbation theory
  (RPT). The blue line for $P_{gg}$ is the prediction of HALOFIT. The overall
  amplitudes of the theoretical curves are scaled by linear bias $b$
  and galaxy-matter cross-correlation coefficient, $r_g$
  (equations~\ref{eq:pgg_real}--\ref{eq:puu_real}). The units of the power
  spectra $P_{gg}, P_{gu}$ and $P_{uu}$ are $(h^{-1} \mathrm{Mpc})^3$,
  $100 \textrm{ km s}^{-1} (h^{-1} \mathrm{Mpc})^3$, and $(100
  \textrm{ km s}^{-1})^2 (h^{-1} \mathrm{Mpc})^3$, respectively. In
  the \textit{bottom} panels, we plot the power spectra divided by the
  linear power spectra. The real-space power spectra are described by
  existing theoretical curves reasonably well.}
\label{fig:power_spectra_real}
\end{figure*}

\subsection{The power spectra in real-space}
\label{sec:realspace-power}
In this section, we present the angle-averaged auto- and cross-power
spectra in real space; we discuss the power spectra in redshift space in
Section~\ref{sec:zspace-power}. We angle-average the power spectra in
the upper-half of Fourier space, which corresponds to an integral
$\int_0^1 d\mu$ in the continuous limit, $V \rightarrow \infty$.  We
average the products of Fourier modes in equally spaced bins with
width $\Delta k= 0.01 \,h\mathrm{Mpc}^{-1}$, and plot the result against the
averages of the magnitudes $k$ in the bins in
Fig.~\ref{fig:power_spectra_real}. The redshift is $z=0$, and the mass
range of the subhaloes is $(10^{11.5}-10^{12}) h^{-1}M_\odot$.  Error
bars indicate the uncertainty due to cosmic variance and shot noise,
\begin{equation}
  \label{eq:delta_p}
  \Delta P_{gg} = (P_{gg} + \bar{n}^{-1})/\sqrt{N_k},
\end{equation}
where $N_k$ is the number of modes in the bin, and $\bar{n} =
5.4\times 10^{-3} (h^{-1} \mathrm{Mpc})^{-3}$ is the subhalo number
density. We do not add the shot noise to the error bars for the
cross-power and the velocity auto-power spectrum: $\Delta P =
P/\sqrt{N_k}$.

We plot the power spectra calculated from perturbation theories as the lines in
Fig.~\ref{fig:power_spectra_real}
--- the linear perturbation theory, the one-loop standard perturbation
theory \citep[see][for a review]{2002PhR...367....1B}, and the one-loop
renormalised perturbation theory \citep[RPT]{2006PhRvD..73f3519C}.  We
also plot the \textsc{HALOFIT} power spectrum
\citep{2003MNRAS.341.1311S, 2012ApJ...761..152T} for the galaxy-galaxy
auto power. We use publicly available codes
\textsc{CAMB}\footnote{\texttt{http://camb.info}}
\citep{2000ApJ...538..473L} for the linear and the \textsc{HALOFIT}
matter power spectra. We calculate the matter-velocity divergence auto-
and cross-power spectra, $P_{mm}$, $P_{m\theta}$, and
$P_{\theta\theta}$, using the Cosmology Routine
Library\footnote{\texttt{http://www.mpa-garching.mpg.de/\~{}komatsu/crl/}}
\citep*{2006ApJ...651..619J} for the standard perturbation theory, and
the \textsc{Copter} package by \citet*{2009PhRvD..80d3531C} for the
one-loop renormalised perturbation
theory.\footnote{\texttt{http://mwhite.berkeley.edu/Copter/}} All
three power spectra are equal to each other in the linear order:
$P_{mm} = P_{m\theta} = P_{\theta\theta}$. We fit the $N$-body result
with two free parameters, the linear galaxy bias $b$ and the galaxy-matter
cross-correlation coefficient $r_g$:
\begin{equation}
  \label{eq:pgg_real}
  P_{gg}(k) = b^2 P_{mm}(k),
\end{equation}
\begin{equation}
  \label{eq:pgu_real}
  P_{gu}(k) = \mathrm{i} aHf \mu b r_g P_{m \theta}(k)/k,
\end{equation}
\begin{equation}
  \label{eq:puu_real}
  P_{uu}(k) = (aHf \mu/k)^2 P_{\theta\theta}(k).
\end{equation}
These functional forms follow from the relation between $u$ and
$\theta$ in equation~(\ref{eq:u_theta}). The galaxy correlation
coefficient can be less than 1 if there is a stochastic bias
\citep{1999ApJ...520...24D}. We calculate the best-fitting bias value
$b=0.92$ by minimising the $\chi^2$ between the simulation and the
HALOFIT power spectra, and find the best-fitting correlation
coefficient value $r_g=0.98$ by similarly fitting between simulation
and RPT cross power $P_{gu}$ for the fixed best-fitting $b$, both in
the range $k \le 0.1 h \mathrm{Mpc}^{-1}$. We assign
equation~(\ref{eq:delta_p}) for the statistical uncertainty in the
power spectra. The figure shows that the one-loop perturbation
theories are in reasonable agreement with the simulation result.

\begin{figure*}
\centering
\includegraphics[width=174mm]{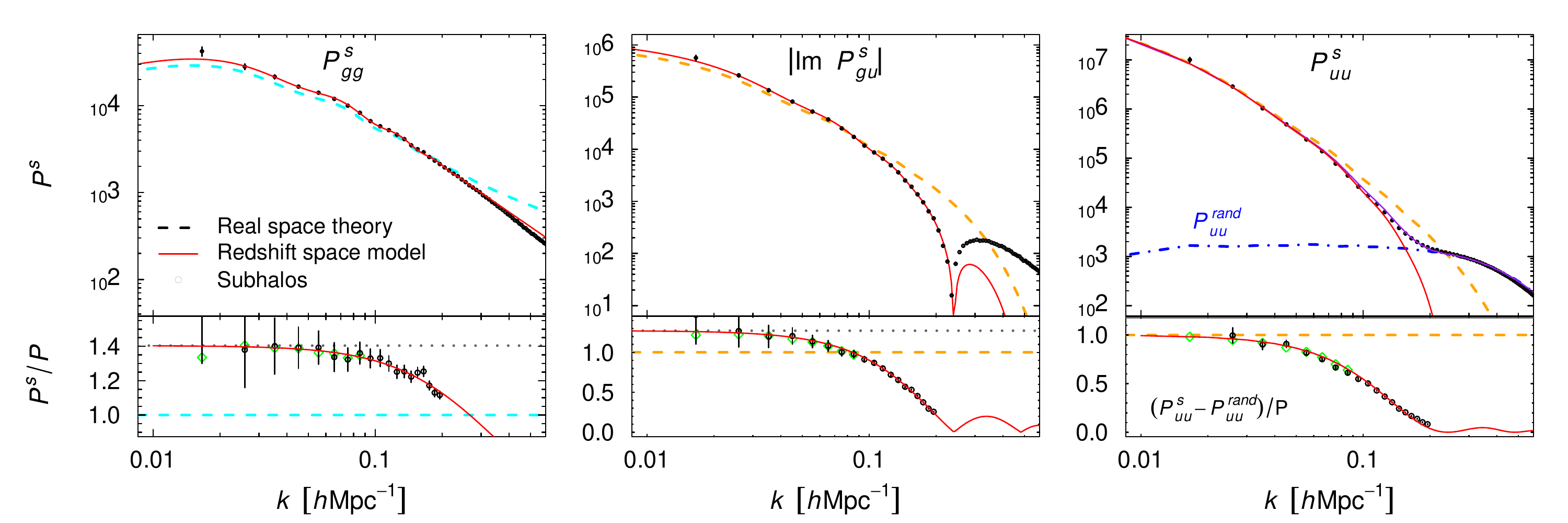}
\caption{The angle-averaged auto- and cross-power spectra of density
  and peculiar velocity in redshift space. The superscript $s$ denotes
  that the power spectra are in redshift space. Our model equations
  (solid red lines) are in good agreement with the subhaloes in the
  GiggleZ simulation (points). The dashed lines are theoretical power
  spectra in real space. \textit{Bottom} panels show the ratio of
  power spectra in redshift space to those in real space. The dotted
  lines are the Kaiser limit. The deviations from the horizontal lines
    show the redshift-space damping.  The circles are the simulation
  data divided by the theoretical real-space curve, while the green
  diamond points at low $k$ are those divided by the simulation power
  spectra in real space.  In the \textit{top right} panel, the random
  velocity component is shown by the blue dash-dotted line, and the
  model including the random term $P_{uu}^\mathrm{rand}$ is
  plotted by the purple solid curve. In the bottom right model, the
  random term is subtracted from the velocity power spectrum. We find
  strong damping in the density-velocity cross-power and velocity
  auto-power spectra.}
\label{fig:power_spectra_model}
\end{figure*}

\subsection{The power spectra in redshift space}
\label{sec:zspace-power}
\subsubsection{The model in redshift space}
\label{sec:vpower_model}
We follow the prescription of modelling density and velocity power
spectra by \citet[][BT04 hereafter]{2004MNRAS.347..255B}, and improve
their model by introducing a new damping term $D_u$ for the
redshift-space distortion of the velocity field. Although
\citetalias{2004MNRAS.347..255B} use the velocity derivative instead
of velocity, both quantities give the same Fisher-matrix forecast, as
we see later.  The galaxy clustering is modelled by a linear galaxy
bias $b = \sqrt{P_{gg}/P_{mm}}$ and a galaxy-mass cross correlation
coefficient $r_g = P_{gm}/\sqrt{P_{gg} P_{mm}}$. The redshift-space
distortion introduces additional perturbation in mass distribution,
described by the Jacobian of the real-to-redshift-space mapping, known
as the Kaiser squashing effect \citep{1987MNRAS.227....1K}.  Unlike
the density or the momentum, the velocity does not have the squashing
term. The same Jacobian term for the density and momentum cancels each
other for the velocity, which is the momentum divided by the density.
We model the redshift-space distortion by the Kaiser factor and a
damping term \citep{1994MNRAS.267.1020P} with the Lorentzian function
\citep{1996MNRAS.282..877B},
\begin{equation}
  D_g(k, \mu)^2 \equiv [ 1 + (k\mu\sigma_g)^2/2 ]^{-1},
\end{equation}
where $\sigma_g$ is a constant related to the pairwise velocity
dispersion. Previous literature does not
include the redshift-space distortions in the velocity field (i.e. sets
$D_u = 1$).

Our model equations for the auto- and cross-power spectra of galaxy
and line-of-sight velocity in redshift space, $P^s_{gg}$, $P^s_{gu}$,
and $P^s_{uu}$ are the following:
\begin{equation}
  \label{eq:pgg_zspace}
  P_{gg}^s(k, \mu) = (1 + 2 r_g \beta \mu^2 + \beta^2 \mu^4) D_g^2 b^2 P_{mm}(k),
\end{equation}
\begin{equation}
  \label{eq:pgu_zspace}
  P_{gu}^s(k, \mu) = 
     \mathrm{i} aHf\mu (r_g + \beta\mu^2) D_g D_u b P_{m\theta}(k)/k,
\end{equation}
\begin{equation}
  \label{eq:puu_zspace}
  P_{uu}^s(k, \mu) =  (aHf\mu/k)^2 D_u^2 P_{\theta \theta}(k),
\end{equation}
where $\beta \equiv f/b$. We expect that these equations hold only on
large scales, $k \la 0.2 \,h\mathrm{Mpc}^{-1}$; non-linearity and
scale-dependent bias become important for higher $k$. As we show
below, we found strong damping in the cross- and velocity 
auto-power spectra. We empirically fit the simulation results by a sinc
function:
\begin{equation}
  \label{eq:du}
  D_u(k) \equiv \sin(k\sigma_u)/(k\sigma_u),
\end{equation}
where the constant $\sigma_u$ is about $13 \,h^{-1}\mathrm{Mpc}$ at
$z=0$ with a small dependence on halo mass (see Table~\ref{table:fitting}).
The model of equation~(\ref{eq:du}) applies to scales $k \le 0.2
\,h\mathrm{Mpc}^{-1}$. In the following, we compare this model with
the simulation in redshift space.

\subsubsection{Fitting the model parameters}
We compute the redshift-space distortion in the simulation by shifting
the line-of-sight coordinates of subhaloes: $x_3 \mapsto x_3 +
u/(aH)$, and measure the power spectra in the same way as in real
space. In Fig.~\ref{fig:power_spectra_model}, we show that our
  model equations for the redshift-space are in good agreement with
  the simulation.  In the upper panels, we show the angle-averaged
auto- and cross-power spectra in redshift space; points are calculated
from subhaloes in the $N$-body simulation, dashed lines are the model
equations in real space (equations
\ref{eq:pgg_real}--\ref{eq:puu_real}), and the red solid lines are our
model equations in redshift space (equations
\ref{eq:pgg_zspace}--\ref{eq:puu_zspace}). The shot noise is
subtracted from the galaxy auto-power spectrum. We use the HALOFIT
model for $P_{mm}$, and RPT for $P_{m\theta}$ and $P_{\theta\theta}$,
respectively. We find the best fitting damping constant $\sigma_g= 5.8
h^{-1} \mathrm{Mpc}$ by fitting the model equation for the
angle-averaged $P^s_{gg}$ to the simulation data with minimum
$\chi^2$ for $k \le 0.2h \mathrm{Mpc}^{-1}$, with fixed values of $b$
and $r_g$ which we have obtained by fitting the real-space power
spectra.  Similarly, we find the best fitting damping constant for the
velocity, $\sigma_u = 13.0 h^{-1}\mathrm{Mpc}$, by fitting the
angle-averaged cross power $P^s_{gu}$ for fixed $b$, $r_g$, and
$\sigma_g$. The angle-average integrals of the model
equations~(\ref{eq:pgg_zspace}--\ref{eq:puu_zspace}) can be performed
analytically with elementary functions.

\subsubsection{The cross-power spectrum}
The cross-power spectrum in redshift space becomes negative at $k=0.23
\,h \mathrm{Mpc}^{-1}$, and returns to positive values at $k=1.06 \,h
\mathrm{Mpc}^{-1}$. This damping in velocity is much larger than that
affecting the galaxy auto-power spectrum; it is a $50$-per cent effect
at $k=0.1\,h\mathrm{Mpc}^{-1}$, and damps to almost zero at
$k=0.2\,h\mathrm{Mpc}^{-1}$. The large damping cannot be explained by
uncorrelated random velocities, in contrast to the damping in the
density field \citep{1994MNRAS.267.1020P}, because such random
displacement would give the same damping for all three auto- and
cross-spectra. The complete correlation between the velocity field and
redshift-space displacement might be the origin of this large
damping. We leave explanations of this strong damping to future
studies. Our empirical damping formula is a good fit for $k \le 0.2\,
h\mathrm{Mpc}^{-1}$, which is sufficient for our Fisher matrix
forecast, but does not capture the shape of negative cross-power spectrum for $k > 0.23 \,h\mathrm{Mpc}^{-1}$. Because
of this oscillating feature, a positive damping function, such as a
Gaussian or Lorentzian, cannot fit the velocity cross-power spectrum
well.

\subsubsection{The velocity auto-power spectrum}
We can explain the velocity auto-power spectrum in redshift space with
the same damping function $D_u$, and an additional \textit{random
  velocity component} $P_{uu}^\mathrm{rand}$, which is plotted with a
blue dash-dotted line in the top-right panel of
Fig.~\ref{fig:power_spectra_model}. We compute $P_{uu}^\mathrm{rand}$
by assigning independent Gaussian random velocities with zero mean and
standard deviation $\sigma_*$ to subhaloes in redshift space. We find
the best fitting value $\sigma_* = 197 \,\textrm{km s}^{-1}$ by
fitting the angle-averaged $P^s_{uu}$ by the sum of
equation~(\ref{eq:puu_zspace}) and $P_{uu}^\mathrm{rand}$, for fixed
$\sigma_u$. We fit for $k\le 0.4 \,h\mathrm{Mpc}^{-1}$ because we can
determine the value of $\sigma_*$ well at high $k$ where the random
component is dominant. We plot the sum of two terms by the solid
purple line, which shows a good fit to the subhalo data. The random
velocity component $P_{uu}^\mathrm{rand}$ is analogous to the shot
noise in galaxy auto-power spectrum, in the sense that they are both
inversely proportional to the galaxy number density, but
$P_{uu}^\mathrm{rand}$ also depends on the virial velocities of the
galaxies. We find that the value of $\sigma_*$ for haloes is smaller
than that for the subhaloes, which is
consistent with our interpretation that $\sigma_*$ is related to
virial motions.

In the bottom panels of Fig.~\ref{fig:power_spectra_model}, we plot
the ratios of power spectra in redshift space to those in real
space. The black circles are $N$-body data divided by the model
equations in real space, while the green diamond points for $k\le 0.1
\,h\mathrm{Mpc}^{-1}$ are the same $N$-body data but divided by the
real-space power spectra calculated from the $N$-body simulation.
The grey dotted lines show the Kaiser limits, which do not have the
damping factors. The data points from the simulation are in good
agreement with our model equations with damping, plotted by the red
lines.

\subsubsection{The angular dependence}
\begin{figure*}
\centering
\includegraphics[width=174mm]{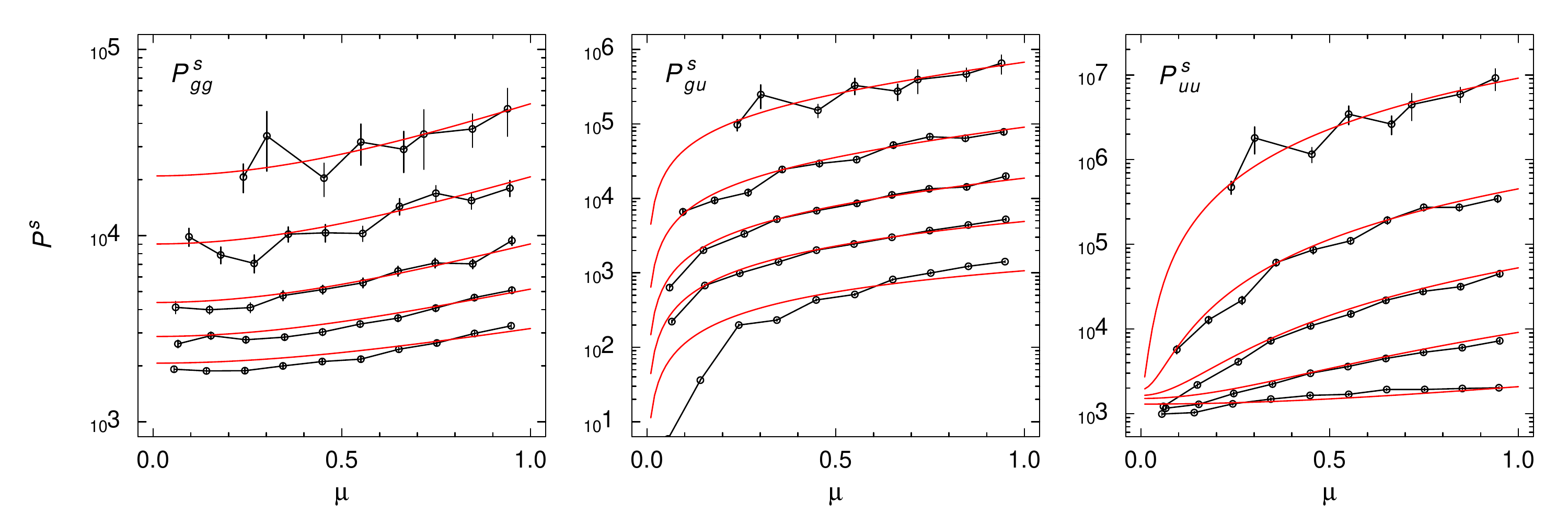}
\caption{The dependence of power spectra on the angle $\mu$ to the
  line-of-sight in redshift-space. The points are calculated from
  subhaloes averaged in bins with widths $\Delta \mu = 0.1$ and
  $\Delta k = 0.01$, at $k= 0.025, 0.065, 0.105, 0.405$ and $0.185
  \,h\mathrm{Mpc}^{-1}$. The red lines are the models with Kaiser
  effect and damping
  (equations~\ref{eq:pgg_zspace}--\ref{eq:puu_zspace}). See
  Fig.~\ref{fig:power_spectra_real} for the units of the power
  spectra.}
\label{fig:model_mu_dependence}
\end{figure*}

In Fig.~\ref{fig:model_mu_dependence}, we plot the auto- and cross-power
spectra as a function of wave vector angle $\mu$ for fixed
$k$. Although $k\mu$ is the natural combination for redshift-space
distortion, we do not see a clear dependence of the velocity damping on
$\mu$. We therefore choose a damping factor that only depends on
$k=|\bmath{k}|$, which is a reasonably good fit for the overall behaviour.\\

Figs.~\ref{fig:power_spectra_model} and \ref{fig:model_mu_dependence}
show that the simple models are consistent with the $N$-body
simulation. The best fitting model parameters depend on the subhalo
sample. We summarise the parameters in Table~\ref{table:fitting}. For
more precise comparisons between simulations and recent models of
galaxy or halo power spectrum in redshift space, see, for example,
\citet{2011PhRvD..84d3526N, 2012ApJ...748...78K, 2012MNRAS.427..327D,
  2012JCAP...11..014O, 2013arXiv1308.6087I}. Although we do not need
accurate models for the Fisher matrix forecast, such accurate models
are important to extract unbiased parameters from data, and increase
the information by extending the range of $k$ that can fit data
without systematic error. Similar work is necessary for the velocity
power spectrum to analyse future peculiar velocity data with high
accuracy. For example, we do not include velocity bias in our model
\citep{2010PhRvD..81b3526D, 2012MNRAS.421.3472E}.

\begin{table}
\caption{The best fitting model parameters for subhaloes in various
  mass ranges, $\log_{10} [M/(1 h^{-1} M_\odot)]=$ 11.5--12.0,
  12.0--12.5, 12.5--13.0, and 13.0--13.5, in addition to the subhalo
  number density $\bar{n}$ in $(h^{-1}\mathrm{Mpc})^{-3}$. The damping
  parameters $\sigma_g$ and $\sigma_u$ are in $h^{-1}\mathrm{Mpc}$,
  and the random velocity parameter $\sigma_*$ is in $\textrm{km
    s}^{-1}$, respectively. All of these parameters depend on the
  subhalo population.}
\label{table:fitting}
\begin{tabular}{@{}ccccccc}
\hline
 Mass range & $\bar{n}$ & $b$ & $r_g$ & $\sigma_g$ & $\sigma_u$ & $\sigma_*$\\
\hline
  11.5 -- 12.0  & $5.4 \times 10^{-3}$  & 0.92 & 0.98 & 5.8 & 13.0 & 197\\
  12.0 -- 12.5  & $2.0 \times 10^{-3}$  & 1.0  & 0.96 & 5.1 & 13.9 & 186\\
  12.5 -- 13.0  & $6.8 \times 10^{-4}$  & 1.2  & 0.94 & 4.3 & 14.4 & 185\\
  13.0 -- 13.5  & $2.1 \times 10^{-4}$  & 1.5  & 0.92 & 3.1 & 15.4 & 195\\
\hline
\end{tabular}
\end{table}

%
% 3. Fisher Information Matrix
%
\section{The Fisher matrix for galaxy number density and velocity}
\label{sec:fisher}
The Fisher information matrix $\mathbfss{F}$ provides the best
possible confidence intervals of unknown parameters, $\theta_i$, such
as $f \sigma_8, \beta$, or $\Omega_m$, under the assumption that the
likelihood function can be approximated by a multivariate Gaussian
about the maximum likelihood. The inverse matrix $\mathbfss{F}^{-1}$
gives the covariance matrix of the parameters $\theta_i$, and $\Delta
\theta_i = (F^{-1})_{ii}$ gives the $1\sigma$ uncertainty in
$\theta_i$, marginalised over all the other parameters. In the context
of large-scale structure, the Fisher matrix forecasts the uncertainties in
cosmological parameters that can be determined from given
observational uncertainties in the power spectrum, which consist of
sample variance of random density fluctuations, and shot noise from
finite number of galaxies \citep{1997PhRvL..79.3806T,
  1998ApJ...499..555T}. Observational error in peculiar velocity
propagates to uncertainties for the velocity power spectrum
\citepalias{2004MNRAS.347..255B}.

We first review the Fisher matrix for the galaxy number density
$\delta_g$ and the line-of-sight velocity $u$ in
Section~\ref{sec:basic_equations}. As \citetalias{2004MNRAS.347..255B}
does not show the detailed derivation of the Fisher matrix with spatially
varying noise term, and the nice derivation by
\citet{2012MNRAS.420.2042A} only focuses on the uncertainty of power
spectra (not of parameters $\theta_i$ in general), we summarise the
mathematical derivation of the Fisher matrix in
Appendix~\ref{sec:fisher_matrix_derivation}, following
\citet{2012MNRAS.420.2042A}. The Fisher matrix for $N$ multiple
tracers (which can be a combination of density and velocity, or
multiple density fields with different galaxy bias) has two 
formulae that are apparently different; one is written as a trace of
$N \times N$ covariance matrices of Gaussian density fields
\citep{2009JCAP...10..007M}, and the other is written as a
bilinear form with the $N(N+1)/2$-dimensional covariance matrix of power spectra
(\citetalias{2004MNRAS.347..255B}, \citealt{2009MNRAS.397.1348W}). We
show in the appendix that these two formulae are algebraically equal to
each other. As a result, the Fisher matrix we use in this paper is
exactly equal to that used by \citetalias{2004MNRAS.347..255B}.

\subsection{The basic equations}
\label{sec:basic_equations}
The Fisher matrix for a multivariate Gaussian random variable with
mean vector $\bmath{\mu}$ and covariance matrix $\mathbfss{C}$ is
\begin{equation}
  \label{eq:fisher_gaussian}
  F_{ij} = 
   \frac{\partial \bmath{\mu}^\textrm{\tiny T}}{\partial \theta_i} 
     \mathbfss{C}^{-1} \frac{\partial \bmath{\mu}}{\partial \theta_j} +
   \frac{1}{2} \mathrm{tr} \left[ \mathbfss{C}^{-1}
    \frac{\partial \mathbfss{C}}{\partial \theta_i} \mathbfss{C}^{-1}
    \frac{\partial \mathbfss{C}}{\partial \theta_j} \right],
\end{equation}
where T represents the vector transpose \citep{1996ApJ...465...34V,
  1997ApJ...480...22T, 1997PhRvL..79.3806T}. 
In a simple case, in which galaxy shot noise $n_g^{-1}$ is spatially
homogeneous, and velocity noise $\sigma_\mathrm{u-noise}$ is also
constant, one can easily derive the following expression for a Fisher
matrix \citep{2009JCAP...10..007M} by applying the formula to real and
imaginary parts of the Fourier modes, $\delta_g^s(\bmath{k})$ and
$u^s(\bmath{k})$, in a periodic box of volume $V$:
\begin{equation}
  \label{eq:fisher_tr_hom}
  F_{ij} = \frac{1}{2} V \int \! \frac{d^3 k}{(2\pi)^3} \,
    \mathrm{tr} \left[
      \tilde{\Sigma}(\bmath{k})^{-1}
      \frac{\partial \tilde{\Sigma}(\bmath{k})}{\partial \theta_i}
      \tilde{\Sigma}(\bmath{k})^{-1}
      \frac{\partial \tilde{\Sigma}(\bmath{k})}{\partial \theta_j}
    \right],
\end{equation}
where the summation over independent $k$ modes is approximated by an
integral $(1/2) V \int d^3x/(2\pi)^3$, and $\tilde{\Sigma}$ is a
matrix of power spectra including noise terms of shot noise and
velocity measurement error,
\begin{equation}
  \tilde{\Sigma} \equiv \left(
    \begin{array}{cc}
      P^s_{gg}(\bmath{k}) + n_g^{-1} & P^s_{gu}(\bmath{k}) \\
      P^s_{ug}(\bmath{k}) & P^s_{uu}(\bmath{k}) + n_u^{-1}\sigma_\mathrm{u-noise}^2
    \end{array}
\right). 
\end{equation}
We allow the number density for shot noise $n_g$ to differ from the number
density of the velocity measurements $n_u$, because galaxies with peculiar
velocity measurements are usually a subset of galaxies with redshift
measurements ($n_u < n_g$); measuring peculiar velocity
requires much higher signal-to-noise ratio in the observations.

In reality, the noise terms vary with distance as the observed
galaxy number density decreases with distance due to the flux
limit of observations, and the peculiar velocity error from standard
candles increases linearly with distance. We assume that the velocity noise
comes from random non-linear motions of rms $\sigma_* \sim 300 \textrm{km
  s}^{-1}$, and observational errors of rms $\sigma_\mathrm{uobs}$, which
originate from the intrinsic scatter in astrophysical relations used as
distance indicators:
\begin{equation}
  \sigma_\mathrm{u-noise}^2 = \sigma_*^2 + \sigma_\mathrm{uobs}^2, \quad 
  \sigma_\mathrm{uobs}(\bmath{x}) = \epsilon H_0^{-1} |\bmath{x}|,
\end{equation}
where the fractional error $\epsilon$ is typically about $8\%$ for
supernovae, and $20\%$ for the Tully-Fisher and the Fundamental Plane distance
indicators. We assume that these noise terms can be determined
directly from observations, and are therefore not a function of
uncertain cosmological parameters $\theta_i$.

It turns out that we can replace the volume $V$ by a volume integral
$\int\! d^3x$ under the `classical approximation'
\citep{1997MNRAS.289..285H, 2012MNRAS.420.2042A}:
\begin{equation}
  \label{eq:fisher_matrix}
  F_{ij} = \frac{1}{2} \int \! \frac{d^3 x d^3 k}{(2\pi)^3} \,
    \mathrm{tr} \left[
      \tilde{\Sigma}(\bmath{k}, \bmath{x})^{-1}
      \frac{\partial \tilde{\Sigma}}{\partial \theta_i}
      \tilde{\Sigma}(\bmath{k}, \bmath{x})^{-1}
      \frac{\partial \tilde{\Sigma}}{\partial \theta_j}
    \right],
\end{equation}
where $\tilde{\Sigma}$ now depends on $\bmath{x}$ through noise terms,
$n_g(\bmath{x})$, $n_u(\bmath{x})$, and
$\sigma_\mathrm{uobs}(\bmath{x})$. This mixture of a wavenumber and a
position seems odd because Fourier transformations of
$\delta(\bmath{x})$ and $u(\bmath{x})$ do not leave $\bmath{x}$ as an
independent variable. We can imagine, however, dividing the volume $V$
into subvolumes where noise terms are approximately constant, Fourier
transforming the fields in each of the subvolume, and obtaining
equation~(\ref{eq:fisher_tr_hom}) in the subvolume. The sum of such
subvolume Fisher matrices gives
equation~(\ref{eq:fisher_matrix}). This process can be justified only
if the wavelength is much smaller then the size of the subvolume, because
(a) Fourier transform may not be possible for wavelength larger than
the size of the subvolume, and (b) the simple summation of `sub-Fisher
matrix' is correct only if the fields are uncorrelated between the
subvolumes. Since long-wavelength modes break these conditions, the
classical approximation is valid for wavelengths smaller than the
scale of noise variation, which is in the order of survey
  size. This original discussion of the classical approximation by
\citet{1997MNRAS.289..285H} for a single density field is generalised
to multiple tracers by \citet{2012MNRAS.420.2042A}. We summarise the
mathematical derivation for equation~(\ref{eq:fisher_matrix}) in
Appendix~\ref{sec:fisher_matrix_derivation}.

For a single field of galaxy density, the Fisher matrix reduces to the
well-known form \citep{1997PhRvL..79.3806T},
\begin{equation}
  \label{eq:fisher_gg_only}
  F_{ij}^\mathrm{gg-only} =
  \frac{1}{2} \int \frac{d^3 x d^3 k}{(2\pi)^3}
    \frac{\partial P^s_{gg}}{\partial \theta_i}
    \frac{\partial P^s_{gg}}{\partial \theta_j}
    \left[ P^s_{gg}(\bmath{k}) + n_g^{-1}(\bmath{x}) \right]^{-1}.
\end{equation}
The Fisher matrix of the peculiar velocity power spectrum has a similar form,
\begin{equation}
  \label{eq:fisher_uu_only}
  F_{ij}^\mathrm{uu-only} =
  \frac{1}{2} \int \frac{d^3 x d^3 k}{(2\pi)^3}
    \frac{\partial P^s_{uu}}{\partial \theta_i}
    \frac{\partial P^s_{uu}}{\partial \theta_j}
    \left[ P^s_{uu} + 
           n_u^{-1} \sigma_\mathrm{uobs}^2 \right]^{-1}.
\end{equation}
We also show results of the Fisher matrix of density-velocity cross power only,
\begin{eqnarray}
  \label{eq:fisher_gu_only}
  F_{ij}^\mathrm{cross-only} \!\!\!\!\!&=&\!\!\!\!\!\!
    \int \frac{d^3 x d^3 k}{(2\pi)^3}
    \frac{\partial P^s_{gu}}{\partial \theta_i}
    \frac{\partial P_{gu}^{s*}}{\partial \theta_j}\nonumber\\
     &&\!\!\!\! \left[ (P^s_{gg} + 
           n_g^{-1})(P^s_{uu} +  n_u^{-1} \sigma_\mathrm{uobs}^2) +
           P_{gu}^{s2}\right]^{-1}.
\end{eqnarray}  
See Appendix~\ref{sec:exaple_covp} for the covariance of the cross-power spectrum.

\subsection{General results}
\label{sec:fisher_results}

Before we forecast cosmological constraints for specific surveys, we
show general results for constant number density $n_g=n_u=\bar{n}$. We
consider how the two-field Fisher matrix for galaxy density and peculiar
velocity (equation~\ref{eq:fisher_matrix}) improves cosmological
constraints compared to those from galaxy redshift only
(equation~\ref{eq:fisher_gg_only}), peculiar velocity only
(equation~\ref{eq:fisher_uu_only}), or cross power only
(equation~\ref{eq:fisher_gu_only}).

The 6-dimensional integral in the Fisher matrix reduces to
a 3-dimensional integral by symmetry. We numerically integrate the
Fisher matrix up to wavenumber $k_\mathrm{max}$ and radius
$r_\mathrm{max}= c H_0 z_\mathrm{max}$, which corresponds to redshift
$z_\mathrm{max}$, where $c=3.0 \times 10^5 \textrm{ km s}^{-1}$ is the
speed of light:
\begin{equation}
  \int d^3 x d^3 k = 
     4\pi \Omega_\mathrm{sky} \int_0^{k_\mathrm{max}}\! dk 
     \int_0^{r_\mathrm{max}}\! dr 
     \int_0^1 \! d\mu,
\end{equation}
where $\Omega_\mathrm{sky}$ is the steradian of the field of view; we
use $\Omega_\mathrm{sky}=4\pi$ in this section, but all results
simply scale as
   $\Delta \theta \propto \Omega_\mathrm{sky}^{-1/2}$.
We integrate up to $k_\mathrm{max}=0.2 \,h\mathrm{Mpc}^{-1}$ and
$z_\mathrm{max}=0.1$ unless otherwise mentioned.

We use equations (\ref{eq:pgg_zspace}--\ref{eq:puu_zspace}) for the
power spectra in redshift space, with $P_{mm}$, $P_{m\theta}$, and
$P_{\theta\theta}$ from the 1-loop renormalised perturbation theory
(RPT). In this section, we set fiducial values, $b=1$, $r_g=1$,
$\sigma_g/\sqrt{2} = 3 \,h^{-1}\mathrm{Mpc}$, $\sigma_u = 13\,
h^{-1}\mathrm{Mpc}$ for the model power spectrum, and set noise terms
using galaxy number density $\bar{n} = 0.01\,(h^{-1}
\mathrm{Mpc})^{-3}$, 20-per cent observational velocity error,
$\epsilon=0.2$, and non-linear random velocity rms $\sigma_* = 300
\textrm{ km s}^{-1}$. We use the $\Lambda$CDM cosmology for the
fiducial value of the growth rate parameter
$f=\Omega_m^{0.55}$ \citep{2005PhRvD..72d3529L}, but we do not
assume this relation between $f$ and $\Omega_m$ in the derivatives in
the Fisher matrix, because we constrain possible deviation of
$f\sigma_8$ from the $\Lambda$CDM cosmology. In the following
subsections, we present the results of the Fisher matrix analysis for different
subsets of cosmological parameters. We summarise the results in
Table~\ref{table:fisher_general}.

\begin{table*}
\begin{minipage}{150mm}
\caption{$1\sigma$ constraints on parameters for constant galaxy
  number density $\bar{n}=0.01 (h^{-1}\mathrm{Mpc})^{-3}$ from
  two-field, galaxy density only (RSD only), and velocity power
  spectrum only.}
\label{table:fisher_general}
\begin{tabular}{llcccccccccc}
& Free parameters & \multicolumn{10}{c}{Fractional uncertainties $\Delta \theta_i / \theta_i$ [per cent]}\\\hline
 &                 & \multicolumn{5}{c}{$k_\mathrm{max}=0.1\,h\mathrm{Mpc}^{-1}$} & \multicolumn{5}{c}{$k_\mathrm{max}=0.2\,h\mathrm{Mpc}^{-1}$}\\
 & $\theta_i$      & $f\sigma_8$ & $\beta$ & $r_g$ & $\sigma_g$ & $\sigma_u$ & $f\sigma_8$ & $\beta$ & $r_g$ & $\sigma_g$ & $\sigma_u$ \\\hline
Two field & $f\sigma_8, \beta \textrm{ (linear)}$ & 2.4 & 2.1 &      &     &     & 1.8 & 1.9 &      &   &   \\
 & $f\sigma_8, \beta$                    & 2.5 & 2.2 &      &     &     & 1.8 & 2.0 &      &   &   \\
 & $f\sigma_8, \beta, r_g$               & 2.5 & 2.2 & 0.30 &     &     & 1.8 & 2.0 & 0.30 &   &   \\
 & $f\sigma_8, \beta, \Omega_b h^2, \Omega_c h^2, h, n_s$ 
                                       & 8.0 & 2.2 &      &     &     & 2.2 & 2.0 &      &   &  \\
 & $f\sigma_8, \beta, \sigma_g, \sigma_u$& 3.3 & 2.9 &      & 65  & 14  & 2.4 & 2.4 &      & 9.6 & 4.6 \\
 & All                                   & 8.5 & 2.9 & 0.30 & 67 & 14   & 2.8 & 2.4 & 0.30 & 11 & 4.7 \\
 & All + Planck prior                    & 3.4 & 2.9 & 0.30 & 65 & 14   & 2.4 & 2.4 & 0.30 & 9.9 & 4.6 
\\\hline
RSD only & $f\sigma_8, \beta $ & 13.8 & 15.8 &      &     &     & 4.9 & 5.6 &      &   & \\
         & $f\sigma_8, \beta, \sigma_g$& 18.7 & 20.2 &      & 85.2  &  & 10.1 & 10.5 &      & 17.0 &  \\
         & $f\sigma_8, \beta, r_g$               & 136 & 134 & 189 &     &     & 48.4 & 47.7 & 67.3 &   &   \\
\hline
Velocity only & $f\sigma_8 $ & 4.0 &  &      &     &     & 3.7 & &      &   & \\
\hline
\end{tabular}
\end{minipage}
\end{table*}

%
% fsigma8 - beta
%
\begin{figure}
\centering \includegraphics[width=84mm]{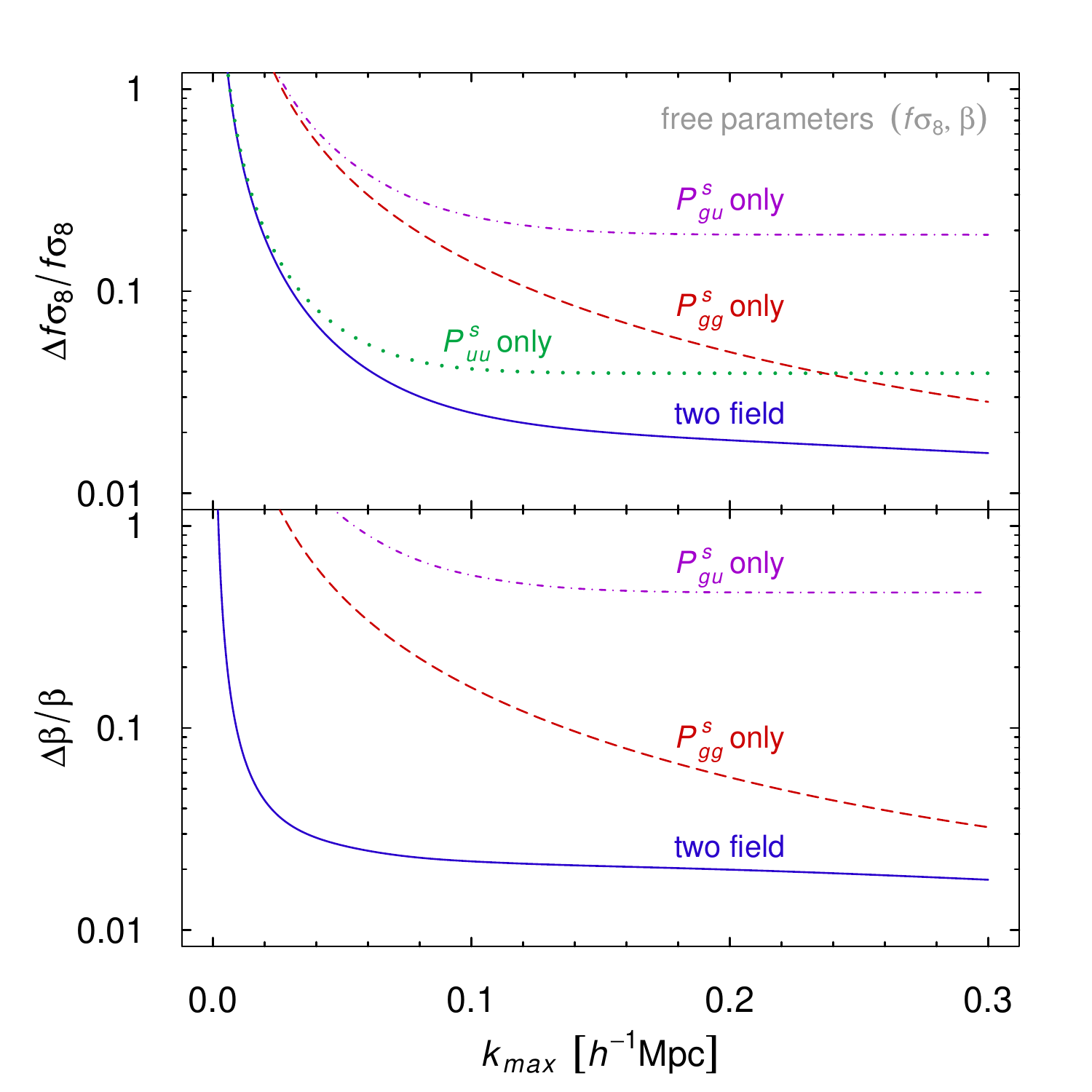}
\caption{Constraints on $f\sigma_8$ (upper panel) and $\beta$ (lower
  panel) as a function of $k_\mathrm{max}$ from galaxy density power
  spectrum only (`$P^s_{gg}$ only', red dashed lines), velocity power
  spectrum only (`$P^s_{uu}$ only', green dotted lines),
  density-velocity cross power only (`$P^s_{gu}$ only', purple
  dash-dotted lines), and from both density and velocity field (`two
  field', blue solid lines). The galaxy number density is fixed to
  $\bar{n}=10^{-2} (h^{-1} \textrm{ Mpc})^{-3}$. Parameters other than
  $f\sigma_8$ or $\beta$ are fixed to their fiducial values, including
  $z_\mathrm{max}=0.1$. The velocity power spectrum alone ($P^s_{uu}$) does
  not constrain $\beta$ because it does not depend on galaxy bias.}
\label{fig:fsigma8_beta_k}
\end{figure}

\subsubsection{Two free parameters: $f\sigma_8$ and $\beta$}
\label{sec:result-two_parameter}
We first show results for two parameters $\bmath{\theta} = (f\sigma_8,
\beta)$, assuming other parameters are exactly known. Because $f$,
$\sigma_8$, and $b$ are completely degenerate, giving an uninvertible
Fisher matrix, we have to select two combinations of three
variables. For the fiducial setup, the constraint on the growth rate
$f\sigma_8$ from the two-field Fisher matrix is $1.8$ per cent, while
the constraint from redshift-space distortion only ($P^s_{gg}$ only)
is $4.9$ per cent. Adding peculiar velocity therefore improves the constraint
by more than a factor of 2.

\begin{figure}
\centering
\includegraphics[width=84mm]{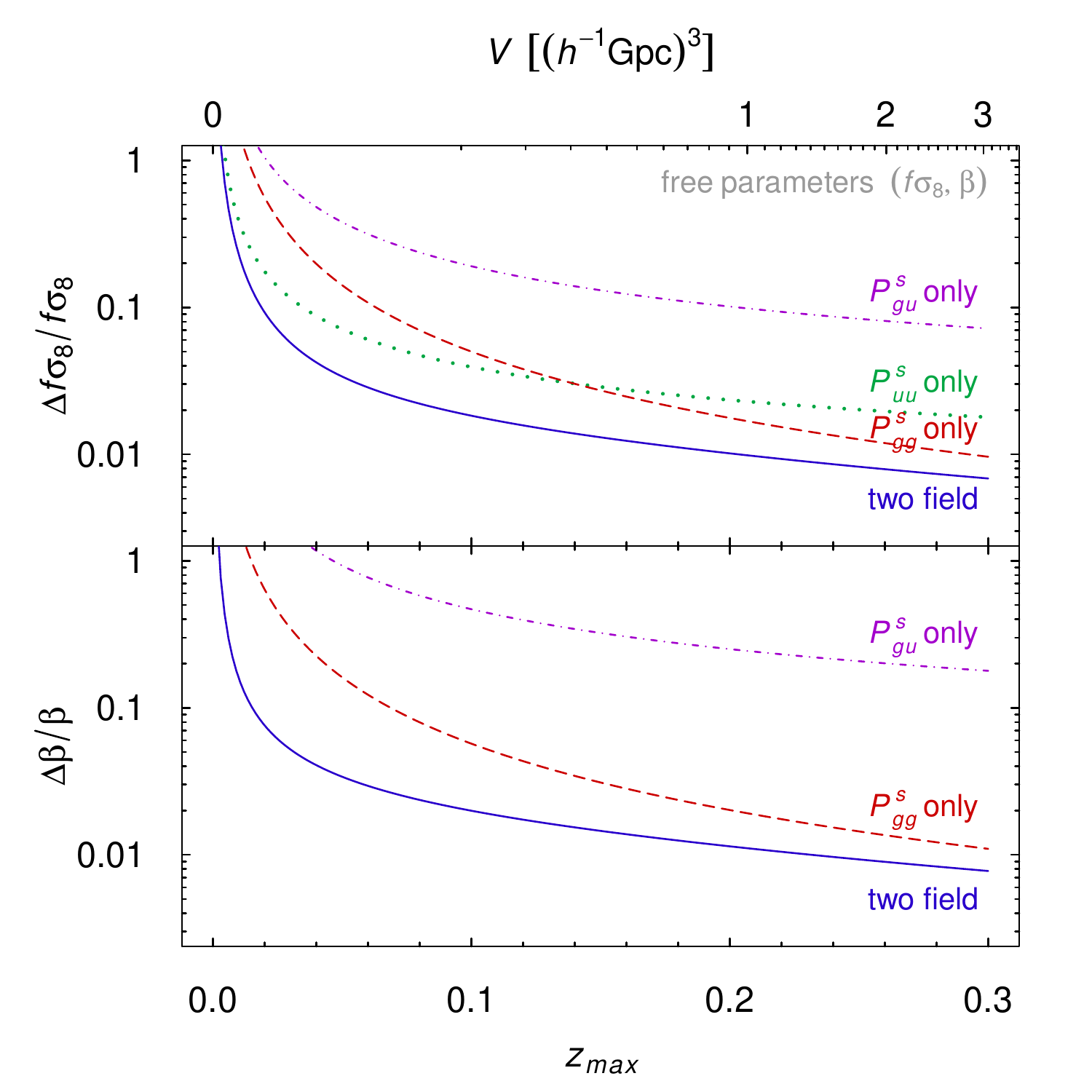}
\caption{Same as Fig.~\ref{fig:fsigma8_beta_k}, but as a function of
  $z_\mathrm{max}$ instead of $k_\mathrm{max}$. $k_\mathrm{max}$ is
  fixed to $0.2\,h\mathrm{Mpc}^{-1}$.}
\label{fig:fsigma8_beta_z}
\end{figure}

In Fig.~\ref{fig:fsigma8_beta_k}, we show the $1\sigma$ constraints on
$f\sigma_8$ and $\beta$ as a function of $k_\mathrm{max}$.  The figure
shows that most of the constraints from velocity come from low $k \le
0.1\,h\mathrm{Mpc}^{-1}$, while constraints from redshift-space
distortion improves at large $k$. The velocity power spectrum damps by
a factor $k^{-2}$ faster than the galaxy power spectrum, and the
measurement error is significant; these two factors make the
signal-to-noise ratio of velocity decline very rapidly as $k$
increases.  In Fig.~\ref{fig:fsigma8_beta_z}, we plot the same
constraints as a function of $z_\mathrm{max}$, for
$k_\mathrm{max}=0.2\,h\mathrm{Mpc}^{-1}$. The constraint from two fields
for $z_\mathrm{max} = 0.1$ are comparable to the constraint from
redshift-space distortion for $z_\mathrm{max} = 0.2$, which has a volume
of $1 (h^{-1} \mathrm{Gpc})^3$.

\begin{figure}
\centering
\includegraphics[width=84mm]{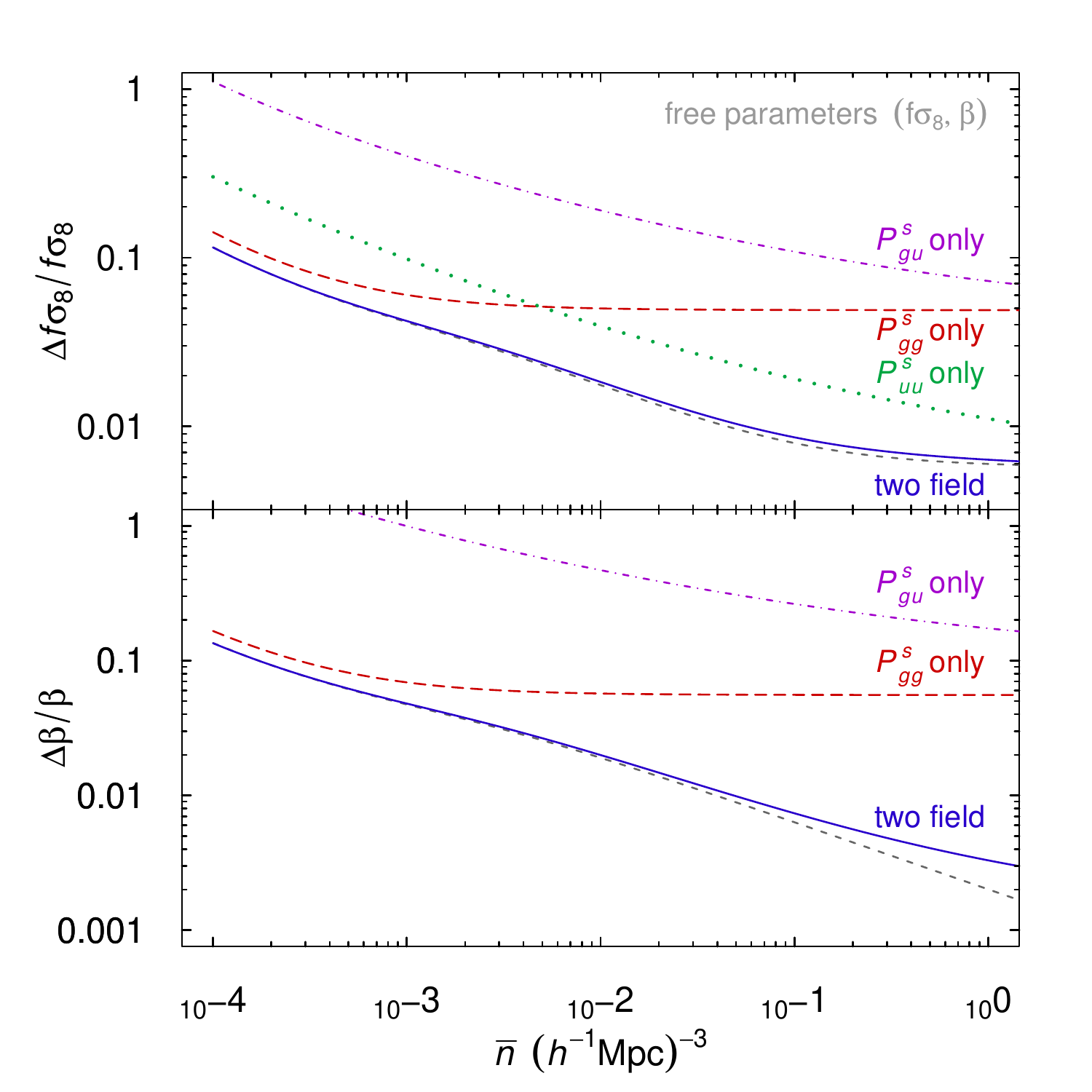}
\caption{Constraints on $f\sigma_8$ (upper panel) and $\beta$ (lower
  panel) as a function of galaxy number density ($\bar{n} = n_g =
  n_u$). See Fig.~\ref{fig:fsigma8_beta_k} for the description of
  lines. The short-dashed lines are results from two fields using
  linear theory; the 1-loop renormalised perturbation theory is used
  for other lines. Constraints on $\beta$ from two fields continue to
  decrease, while the constraint from RSD only is limited by cosmic
  variance.}
\label{fig:fsigma8_beta_ngal}
\end{figure}

In Fig.~\ref{fig:fsigma8_beta_ngal}, we plot the constraints as a
function of galaxy number density $\bar{n}$, for
$k_\mathrm{max}=0.2\,h\mathrm{Mpc}^{-1}$ and
$z_\mathrm{max}=0.1$. While constraints from redshift-space distortion
alone reach the cosmic variance limit at $\bar{n} \sim 10^{-4}
(h^{-1}\mathrm{Mpc})^3$, the constraints from two fields improve
further with galaxy number density. Although the constraint on $\beta$
is not limited by cosmic variance (at linear order with $r_g=1$), the
constraint on $f\sigma_8$ is limited by the cosmic variance on the
clustering amplitude of galaxies, $b\sigma_8$. The cosmic variance
limit of $f\sigma_8$, however, is at the sub-per cent level because
$b\sigma_8$ can be measured very precisely. The bottom panel
illustrates that the constraint on $\beta$ using both density and
velocity is not limited by cosmic variance; the constraint continues
to improve as number density increases, as the large number reduces
the measurement error of velocity on average. The cancellation of
cosmic variance is not perfect for nonlinear power spectra because
$P_{mm}$, $P_{m\theta}$ and $P_{\theta\theta}$ are not exactly equal
to each other. However, the difference between linear and nonlinear
power spectra affects the Fisher matrix results only at very high
number density, $\bar{n} \ga 0.1 (h^{-1}\mathrm{Mpc})^{-3}$, probably
because signal to noise is otherwise not good enough anyway for $k \ga
0.1 h\mathrm{Mpc}^{-1}$, where the non-linearity makes the
difference. In the figure, we plot the constraints using the linear
power spectrum with the blue short dashed lines.

\subsubsection{Three free parameters: $f\sigma_8$, $\beta$, and $r_g$}
\label{sec:result-three_parameter}
From galaxy density alone, the growth rate $f\sigma_8$ and galaxy
correlation coefficient $r_g$ are highly
degenerate. \citetalias{2004MNRAS.347..255B} pointed out that peculiar
velocity breaks this degeneracy and constrains $r_g$ extremely well.
Our result confirms this; the constraint on $f\sigma_8$ from the redshift
survey weakens from $5$ per cent to 48 per cent, compared to
the 2-parameter case (Section~\ref{sec:result-two_parameter}), while
two-field data constrains $r_g$ to $0.3$ per cent, and $f\sigma_8$ to
the same precision as the 2-parameter case. Peculiar velocity surveys
can constrain growth rates, $f\sigma_8$ and $\beta$, equally well even
if we add $r_g$ as a free parameter.

\subsubsection{Four free parameters: $f\sigma_8$, $\beta$, $\sigma_g$ and 
$\sigma_u$}
\label{sec:result-four_parameter}
Because the damping factor of the galaxy power spectrum, $\sigma_g$,
is affected by complicated non-linear pairwise velocity
\citep[e.g.,][]{2004PhRvD..70h3007S}, which depends on the galaxy
population, $\sigma_g$ is often treated as a nuisance parameter fitted
against data. For the velocity damping factor, $\sigma_u$, we do not
yet have a theoretical model. Without knowing how it depends on
cosmological parameters, we have to treat it as a free parameter as
well. We investigate the effects of treating these damping factors as
free parameters in this section. Because we know the order of
magnitude of these parameters and know that they are positive, we add
$100$-per cent priors to the Fisher matrix:
\begin{equation}
  F_{\sigma_g\sigma_g}^\mathrm{\sigma prior} = \sigma_g^{-2}, \quad
  F_{\sigma_u\sigma_u}^\mathrm{\sigma prior} = \sigma_u^{-2}.
\end{equation}
The constraints on $f\sigma_8$ and $\beta$ weaken by about 20 to 30
per cent, from $1.8\%$ to $2.4\%$ on $f\sigma_8$, and from $2.0\%$ to
$2.4\%$ on $\beta$, respectively. The constraint from
redshift-distortion alone also weakens from $5\%$ to $10\%$. We
conclude that uncertainty in the damping parameter has a moderate, but
not severe, effect on the forecast constraints.

\subsubsection{Free cosmological parameters}
Finally, we vary cosmological parameters, cold dark matter density
$\Omega_c h^2$, baryon density $\Omega_b h^2$, Hubble constant $h$,
and spectral index $n_s$ in addition to $f\sigma_8$ and $\beta$. We
take the derivative with respect to cosmological parameters
numerically by generating power spectra with cosmological parameters
changed by $\pm 1$ per cent,
\begin{equation}
  \frac{\partial P}{\partial \theta_i} \approx
  \frac{P(\theta_i + \Delta \theta_i) - P(\theta_i - \Delta \theta_i)}
       {2\Delta \theta_i},
\end{equation}
where $\Delta \theta_i = 0.01\theta_i$. The constraint on $\beta$ is
unaffected, because the relation between $\delta_g$ and $u$ only
depends on $\beta$, not on other cosmological parameters in the linear
order. The constraint on $f\sigma_8$ weakens from $1.8$ to $2.2$ per
cent.

Since cosmological parameters are well constrained by the cosmic
microwave background (CMB), we add the prior expected from the
\textit{Planck} observation \citep{2013arXiv1303.5076P}. We use the
forecast for the full \textit{Planck} mission by
\citet{2006JCAP...10..013P}; we calculate the covariance matrix of
$\Omega_c h^2$, $\Omega_b h^2$, $h$ and $n_s$, marginalised over the
other parameters, using their publicly available Markov chain Monte
Carlo
data.\footnote{\texttt{lesgourg.web.cern.ch/lesgourg/codes/chains\_0606227.html}}
We add the inverse of the covariance matrix to the Fisher matrix as an
independent prior from \textit{Planck}. We do not add a prior on $f$
or $\sigma_8$ from the CMB, because model-dependent extrapolation to
$z=0$ is necessary for such constraints. The \textit{Planck} priors
marginalised for each parameter are $\Delta \Omega_b h^2 = 0.00022$,
$\Delta \Omega_c h^2 = 0.0024$, $\Delta h = 0.017$, and $\Delta n_s =
0.0074$.

After adding the Planck prior, the constraints on $f\sigma_8$ and
$\beta$ recover the 2-parameter constraint. We also vary all 9
parameters, $\bmath{\theta} = (f\sigma_8, \beta, r_g, \sigma_g,
\sigma_u, \Omega_c h^2, \Omega_b h^2, h, n_s)$, with the Planck
prior. The result is same as the 4-parameter constraint with
$f\sigma_8$, $\beta$, $\sigma_g$, and $\sigma_u$. With the precise
measurement from the CMB, the shape of the power spectrum is no longer a
source of uncertainty in the growth rate. \\

We have presented the results of the two-field Fisher matrix analysis
for galaxy density and peculiar velocity, comparing with those for a
single field with density or velocity only. Peculiar velocity
measurements improve the measurements for more than a factor of 2
compared to density alone for $\bar{n}=10^{-2} \,(h^{-1}
\mathrm{Mpc})^{-3}$, and improve even more as we increase the number
density, without the cosmic variance limit. The nonlinear power
spectrum with RPT does not alter the Fisher matrix results
significantly compared to the linear power spectrum. The uncertainty
in the redshift-space damping parameters, $\sigma_g$ and $\sigma_u$,
degrade the constraints by $20-30$ per cent, which can be improved by
future theoretical work.

%
% 4. Forecasting future peculiar velocity survey
%
\section{Forecasting future peculiar velocity surveys}
\label{sec:forecast}
\begin{figure}
\centering \includegraphics[width=84mm]{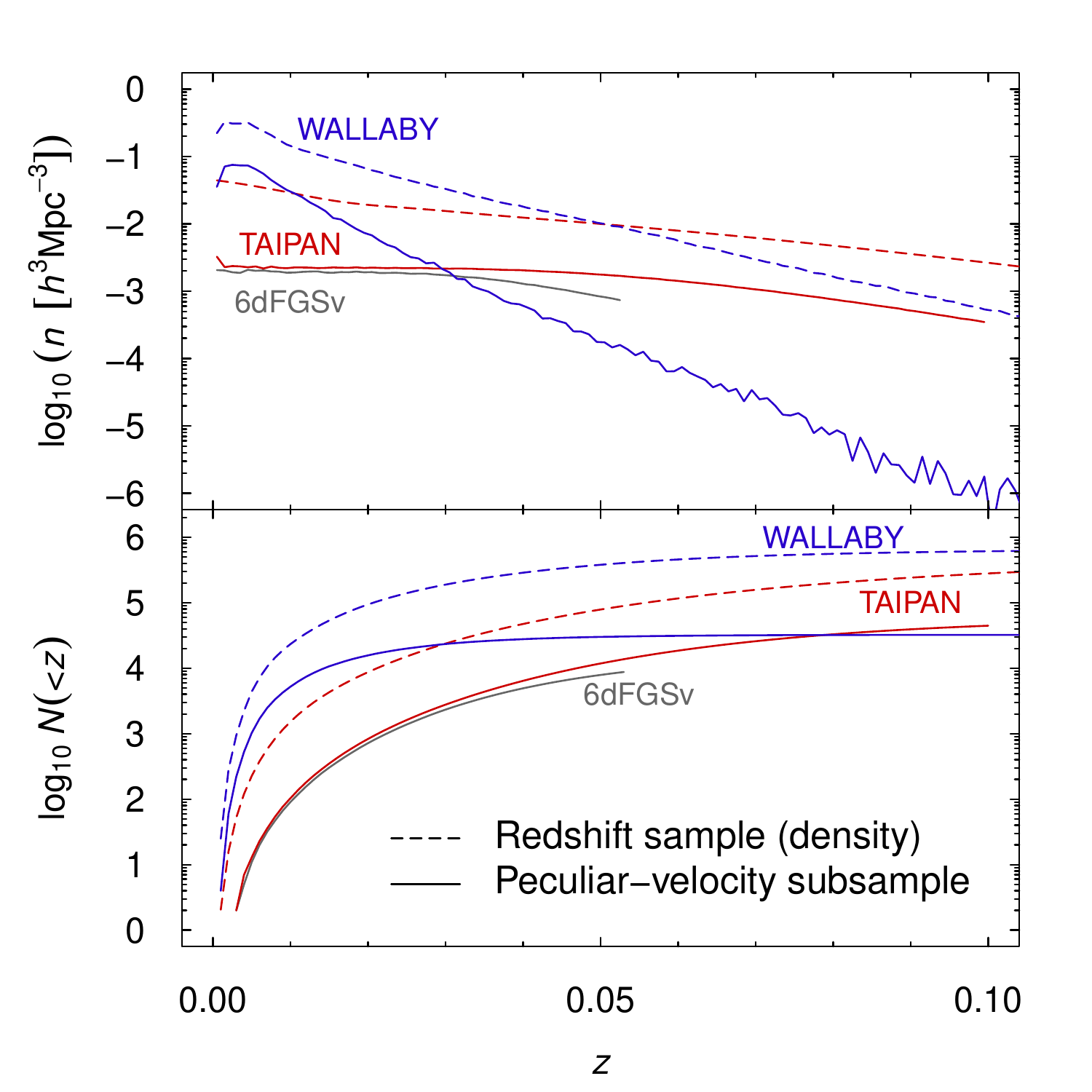}
\caption{Number density (\textit{upper} panel) and cumulative number
  (\textit{lower} panel) of galaxies for the 6dF Galaxy Survey
  velocity subsample (6dFGSv, grey line), TAIPAN survey (red lines),
  and WALLABY+WNSHS survey (blue lines). Dashed lines are for the
  redshift (density only) samples, and solid lines are for the
  peculiar velocity subsamples. }
\label{fig:survey_n}
\end{figure}

We apply our Fisher matrix of galaxy density and peculiar velocity to
existing and future peculiar velocity surveys. In
Section~\ref{sec:6df}, we first review the measurements from the
existing 6dF Galaxy Survey, and the Fisher matrix forecast by
\citetalias{2004MNRAS.347..255B} for the 6dF survey, and compare them
with our calculations. We then present the forecasts for the future
surveys in Sections~\ref{sec:taipan}--\ref{sec:wallaby}. We use
distance-dependent galaxy numbers, galaxy bias $b$, and sky coverage
$\Omega_\mathrm{sky}$ expected for each of the surveys, which we
describe in the following sections. Other parameters in the Fisher
matrix are the same as those in Section~\ref{sec:fisher_results}. In
Fig.~\ref{fig:survey_n}, we plot the expected galaxy number densities
with redshift measurement, $n_g$, and with additional peculiar
velocity measurement, $n_u$. We summarise the results in
Table~\ref{table:forecast}.

\begin{table*}
\begin{minipage}{160mm}
\caption{A Fisher matrix forecast for fractional uncertainties of
  parameters $\Delta \theta_i/\theta_i$, including constraints on
  growth factors $f\sigma_8$ and $\beta$ from the two-field Fisher matrix
  of galaxy density and peculiar velocity, galaxy density only (`RSD
  only'), and velocity only (`$P_{uu}$`). The upper block of numbers are
  for $k_\mathrm{max}=0.1 \,h\mathrm{Mpc}^{-1}$, and the lower block is
  for $k_\mathrm{max}=0.2 \,h\mathrm{Mpc}^{-1}$.}
\label{table:forecast}
\begin{tabular}{llcccccccccc}
 Survey & Free parameters & \multicolumn{10}{c}{Fractional uncertainties $\Delta \theta_i/\theta_i$ [per cent]}\\\hline
                 &  & \multicolumn{5}{c}{Two fields} & \multicolumn{4}{c}{RSD only} & \multicolumn{1}{c}{$P_{uu}$}\\
                 & $\theta_i$      & $f\sigma_8$ & $\beta$ & $r_g$ & $\sigma_g$ & $\sigma_u$ & $f\sigma_8$ & $\beta$ & $r_g$ & $\sigma_g$ & $f\sigma_8$ \\\hline
6dFGSv   & $f\sigma_8, \beta$ & 15 & 16 & & & & & & & & 25\\
         & $f\sigma_8, \beta, r_g$ & 15 & 16 & 4.2 & & & & &\\
         & $f\sigma_8, \beta, \sigma_g, \sigma_u$ & 18 & 19 & & 96 & 66 & \\
TAIPAN   & $f\sigma_8, \beta$ & 7.1 & 7.4 & & & & 28 & 31 & & & 14\\
         & $f\sigma_8, \beta, r_g$ & 7.1 & 7.4 & 0.87 & & & 370 & 360 & 47\\
         & $f\sigma_8, \beta, \sigma_g, \sigma_u$ & 8.9 & 9.1 & & 33 & 36 & 92\\
WALLABY+WNSHS
         & $f\sigma_8, \beta$ (linear) & 3.5 & 3.8 & & & & 11 & 13 & & & 13\\         
         & $f\sigma_8, \beta$ & 4.3 & 4.8 & & & & 11 & 13 & & & 13\\
         & $f\sigma_8, \beta, r_g$ & 4.4 & 4.8 & 0.29 & & & 80 & 78 & 120\\
         & $f\sigma_8, \beta, \sigma_g, \sigma_u$ & 6.0 & 6.4 & & 70 & 23 & 14 & 16 & 85 & 79\\
         & All + \textit{Planck} prior & 6.1 & 6.4 & 0.29 & 70 & 23 & 81 & 80 & 120\\
\multicolumn{12}{r}{[$k_\mathrm{max}=0.1 \, h\mathrm{Mpc}^{-1}$]}\\
\hline \hline
6dFGSv   & $f\sigma_8, \beta$ & 12 & 13 & & & & & & & & 24\\
         & $f\sigma_8, \beta, r_g$ & 12 & 13 & 4.2 & & & & &\\
         & $f\sigma_8, \beta, \sigma_g, \sigma_u$ & 15 & 16 &  & 43 & 66 & \\
TAIPAN   & $f\sigma_8, \beta$ & 5.6 & 6.2 & & & & 10 & 11 & & &14\\
         & $f\sigma_8, \beta, r_g$ & 5.6 & 6.1 & 0.87 & & & 130 & 130 & 170\\
         & $f\sigma_8, \beta, \sigma_g, \sigma_u$, & 7.5 & 7.6 & & 16 & 17 & 33 & 37 & 93\\
WALLABY+WNSHS
         & $f\sigma_8, \beta$ (linear) & 2.6 & 3.0 & & & & 4.3 & 5.2 & & &11\\
         & $f\sigma_8, \beta$ & 3.0 & 3.6 & & & & 4.3 & 5.2 & & &11\\
         & $f\sigma_8, \beta, r_g$ & 3.0 & 3.6 & 0.29 & & & 32 & 31 & 50\\
         & $f\sigma_8, \beta, \sigma_g, \sigma_u$, & 4.5 & 4.7 & & 13 & 7.0 & 8.3 & 8.9 & 18\\
         & All + \textit{Planck} prior & 4.6 & 4.8 & 0.29 & 14 & 7.0 & 34 & 33 & 50 & 20\\
\multicolumn{12}{r}{[$k_\mathrm{max}=0.2 \, h\mathrm{Mpc}^{-1}$]}
\\\hline
\end{tabular}
\end{minipage}
\end{table*}

\subsection{6dF Galaxy Survey}
\label{sec:6df}
The 6dF Galaxy Survey (6dFGS) is a low redshift survey of early type galaxies
out to $z \la 0.15$, covering 17 046 deg$^2$ in the southern sky, and
containing 136 304 redshifts \citep{2004MNRAS.355..747J,
  2009MNRAS.399..683J}. The velocity subsample (6dFGSv) contains 8896
galaxies in the redshift range $z \le 0.05$, with peculiar velocities
measured through the Fundamental Plane relation
\citep{2012MNRAS.427..245M}.  The redshift survey measured the growth
rate through the redshift-space distortions with $13$ per cent
precision at effective redshift of 0.067 \citep[$f\sigma_8 = 0.423 \pm
  0.053$]{2012MNRAS.423.3430B}. Their Fisher matrix calculation gives
constraints on $f\sigma_8$ of $23$ per cent for
$k_\mathrm{max}=0.1\,h\mathrm{Mpc}^{-1}$, and 8.3 per cent for
$k_\mathrm{max} = 0.2\,h\mathrm{Mpc}^{-1}$, consistent with their actual
analysis (all values for wavenumbers $k$ are in units of
$h\mathrm{Mpc}^{-1}$, hereafter).  The velocity subsample combined
with the reconstructed velocity field from a full-sky density field
measures $\beta$ with about 25 per cent precision (Magoulas et al. in
preparation). 

In order to determine the redshift distribution of galaxies, we use
125 random mocks of the 6dFGS velocity subsample, each of which
contains 8986 galaxies (We recently removed 90 objects from the
subsample, which have problems with photometry or spectroscopy. The
number we use here is that before the removal.) We require the J band
luminosity $J \le 13.65$, and velocity dispersion to be larger than
$116\textrm{ km s}^{-1}$ \citep[see,][for the details about the mock
  sample]{2012MNRAS.427..245M}. Velocity subsample is limited to
$z_\mathrm{max}=0.055$ because the velocity dispersion measurements
for the Fundamental-Plane relation become dominated by systematics
beyond that redshift with the 6dF spectrograph. Because the Fisher
  matrix analysis for the redshift sample is already discussed in
  \citet{2012MNRAS.423.3430B}, we only consider the velocity subsample in
  this paper, $n_g = n_u$. The field of view is the southern half of
the sky, excluding the Galactic plane (Galactic latitude $|b| < 10$
deg), which is $\Omega_\mathrm{sky}=1.65\pi$ steradian.  We set galaxy
bias to $b=1.4$ \citep{2012MNRAS.423.3430B}, and fractional velocity
measurement error to $\epsilon=0.25$.

Our Fisher matrix calculation with two free parameters ($f\sigma_8$
and $\beta$) gives $16$ per cent for $k_\mathrm{max}=0.1$, and $13$
per cent for $k_\mathrm{max}=0.2$, respectively, on the uncertainty of
$\beta$. This is in the same order as the peculiar velocity analysis
by Magoulas et al. (in prep), but we cannot compare our Fisher matrix
results directly with their analysis, because our Fisher matrix
results are the combined constraints from redshift-space distortions
and the velocity measurements, while the velocity-velocity analysis
uses the density to construct the model velocity, but is not combined
with the redshift-space-distortion analysis for the density.

\citetalias{2004MNRAS.347..255B}, on the other hand, predicted
a precision of $5$ per cent for the 6dFGS from redshift-space distortion
alone, and $2$ per cent from the two-field Fisher matrix for density
and velocity. Both of these forecasts are much smaller than the
results of the 6dF survey, including the Fisher matrix calculation in
\citet{2012MNRAS.423.3430B}. Details, such as cosmological parameters
or power spectrum model, cannot explain the difference. One
possibility is that they might have used an order of magnitude larger
number density, even though their estimates of the total galaxy
number, about $10^5$ redshifts and about $15 000$ velocity
measurements, are roughly correct; their equation~(29) relates total
number for galaxies $N_g$ to number density (number of galaxies per
unit volume), but the actual integral of the equation gives $4\pi N_g$ in
total, which means that the number density could be $4\pi$ larger than
it should be. If this were true, it would explain why their forecast
is closer to our forecast for WALLABY rather than for 6dFGS.

\subsection{TAIPAN survey}
\label{sec:taipan}
Transforming Astronomical Imaging surveys through Polychromatic
Analysis of Nebulae (TAIPAN) survey is a planned future successor of
the 6dF Galaxy Survey using the UK Schmidt Telescope with upgraded
fibres. The new spectrograph improves the velocity dispersion
measurements, which allows us to extend the upper limit of the
velocity subsample from 0.055 for 6dFGSv to 0.1, and to increase the
number density by decreasing the lower bound of the velocity dispersion from
$116 \textrm{ km s}^{-1}$ for 6dFGSv to $70 \textrm{ km s}^{-1}$.

We generated 125 mock Fundamental-Plane galaxies for TAIPAN, similar
to those for the 6dFGSv, to estimate the number of observed
galaxies. We assume peculiar velocity measurements are available for
J-band magnitude brighter than 15.15, and velocity dispersion larger
than $70 \textrm{ km s}^{-1}$ up to $z=0.1$. We estimate the total
number of velocity sample to be about 45 000. The number of redshifts
in the sample increases by about a factor of 4 compared to 6dFGS, out to $z
\sim 0.2$, assuming an r-band magnitude limit of 17. We assume a galaxy
bias $b=1.4$, same as for 6dFGS, and a fractional velocity error
$\epsilon=0.2$.

The TAIPAN survey can constrain $f\sigma_8$ to 7.1 per cent
($k_\mathrm{max}=0.1$) or 5.6 per cent ($k_\mathrm{max}=0.2$) if the
damping constants, $\sigma_g$ and $\sigma_u$, are known, or 8.9 per
cent ($k_\mathrm{max}=0.1$) to 7.5 per cent ($k_\mathrm{max}=0.2$) if
they are unknown nuisance parameters. Therefore the constraints from the TAIPAN
survey are expected to be a factor of 2 better than those from the
6dFGSv. We will also show in Section~\ref{sec:k-dependence} that
TAIPAN will be able to put interesting constraints on $k$-dependent
growth rates.

\subsection{The WALLABY and the WNSHS surveys}
\label{sec:wallaby}
WALLABY\footnote{http://www.atnf.csiro.au/research/WALLABY} (Widefield
ASKAP L-band Legacy All-sky Blind surveY) is a planned H\,{\sc i}
survey with the Australian SKA Pathfinder (ASKAP), covering $3\pi$
steradian of sky \citep[][and references therein]{2008ExA....22..151J,
  askap, 2012MNRAS.426.3385D}. A similar survey,
The Westerbork Northern Sky H\,{\sc i} Survey (WNSHS),\footnote{\texttt{http://www.astron.nl/\~{}jozsa/wnshs}} is
planned in the other $\pi$ steradian in the northern hemisphere using
APERTIF on the Westerbork Synthesis Radio Telescope. For simplicity,
we assume that both surveys will have the predicted WALLABY rms of 1.6
mJy in a channel of width $3.9 \textrm{ km s}^{-1}$. WNSHS may be
somewhat deeper in practice. This results in a 5-$\sigma$
(velocity-integrated) redshift catalogue of $\sim 0.8$ million objects
\citep{2012MNRAS.426.3385D}. For Tully-Fisher velocities, additional
constraints of width $> 80 \textrm{ km s}^{-1}$, inclination $> 30
\textrm{ deg}$ and 3-$\sigma$ per channel result in a reduced mock
catalogue of $\sim 32 000$ galaxies.

We assume galaxy bias $b=0.7$ following \citet{2012MNRAS.423.3430B},
which is based on measurements from the H\,{\sc i} Parkes All-Sky
Survey \citep{2007MNRAS.378..301B}. The fractional velocity error is
set to $\epsilon=0.2$. Our Fisher matrix forecast for this all-sky
H\,{\sc i} is 3.0 per cent for $f\sigma_8$, and 3.6 per cent for
$\beta$, for $k_\mathrm{max}=0.2$, if we know the values of damping
constants $\sigma_g$ and $\sigma_u$. The non-linear effect slightly
weakens the constraint; the forecasts using the linear power spectra
are 2.6 per cent for $f\sigma_8$ and $3.0$ per cent for $\beta$,
respectively. If we marginalise over damping constants, the
constraints degrade by $30-50$ per cent. Compared to redshift
measurements alone, adding peculiar velocity data reduces the
uncertainties by about 40 per cent. The galaxy correlation coefficient
$r_g$ can be constrained to $0.3$ per cent.

\citet{2012MNRAS.423.3430B} have forecast constraints on the growth
rate $f\sigma_8$ from redshift-space distortion alone: $10.5$ per cent
for $k=0.1$, and $3.9$ per cent for $k=0.2$, respectively. Our
forecasts are consistent with these results, although slightly larger for
$k=0.2$. They also reported that the multiple-tracer method
\citep{2009JCAP...10..007M} between early type galaxies from TAIPAN
survey and gas-rich galaxies from the WALLABY survey in the overlap
volume does not improve the constraints, giving almost the same
constraint as WALLABY+WNSHS only. This is because the number
density for TAIPAN is not large enough to make the multiple
biased-tracer method effective. The method works best when both of the
galaxy populations have high number densities, and simultaneously have
a large difference in their bias, which is a difficult
condition to satisfy. The advantage of the peculiar velocity survey is
that two tracers (density and velocity) are available with high
densities, only limited by the condition that the Tully-Fisher
relation holds.

\begin{figure}
\centering \includegraphics[width=84mm]{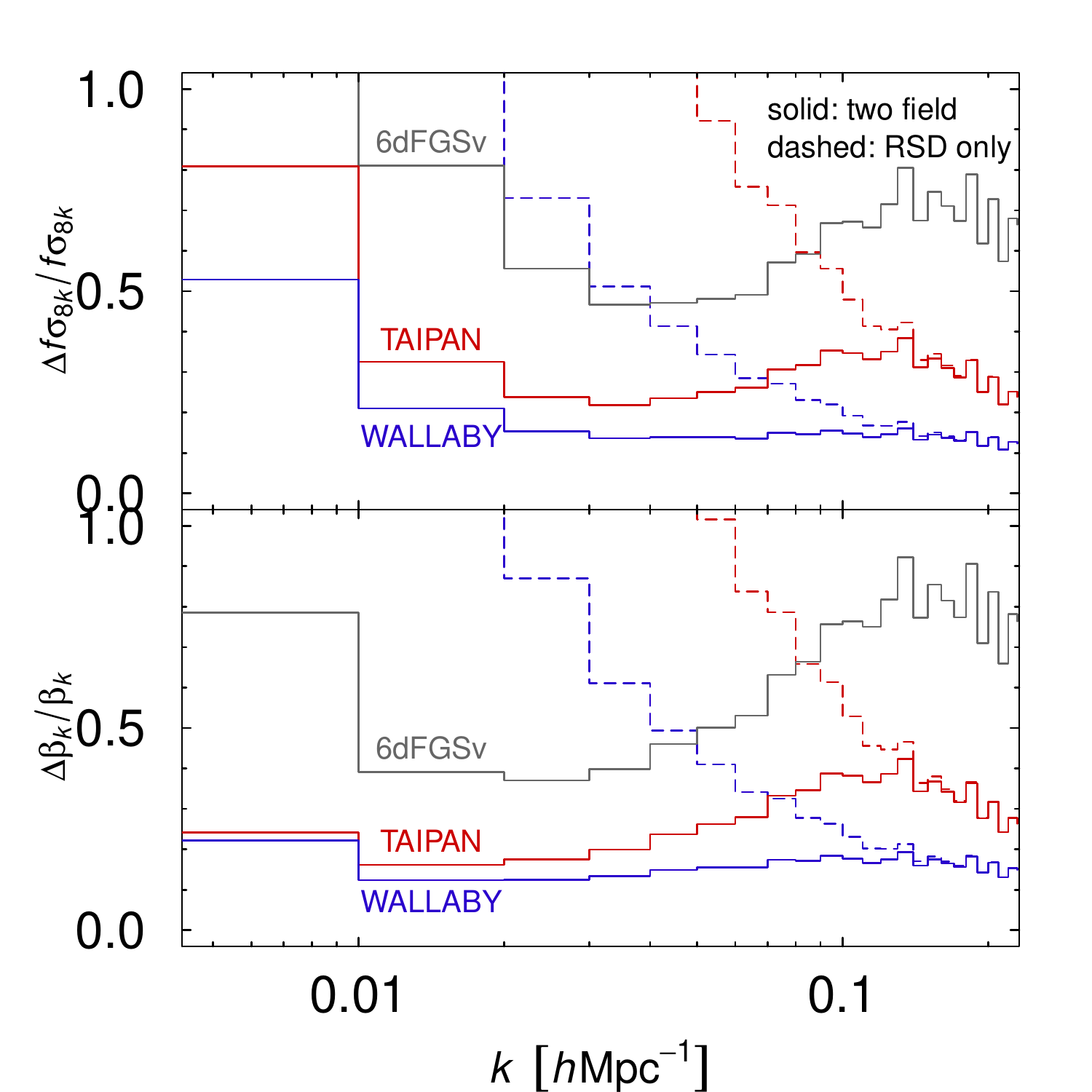}
\caption{Constraints on $f\sigma_8$ and $\beta$ as a function of
  wavenumber $k$, in bins of widths $\Delta k=
  0.01\,h\mathrm{Mpc}^{-1}$. See Fig.~\ref{fig:survey_n} for line
  types and colours. Peculiar velocity surveys improve the constraints
  in $k$ bins for $k \le 0.1\,h\mathrm{Mpc}^{-1}$, because they are not
  limited by the cosmic variance. $k$-dependence constrains modified
  theories of gravity that have scale-dependent growth rates.}
\label{fig:k-dependence}
\end{figure}

\subsection{Constraining $k$ dependence}
\label{sec:k-dependence}
Measuring growth rates, $f\sigma_8$ or $\beta$ on different scales, as
a function of wavenumber $k$ for example, is an independent test of
General Relativity on cosmological scales. General Relativity predicts
that these growth rates are functions of time only, independent of the
wavenumber, but other theories of gravity can have $k$-dependent
growth rates. The growth rate $\beta$ can also be scale-dependent at
large scales if non-Gaussian initial conditions introduce
scale-dependent bias.  In Fig.~\ref{fig:k-dependence}, we show
constraints on the growth rates in bins of width $\Delta k =
0.01\,h\mathrm{Mpc}^{-1}$. We integrate the Fisher matrix with two
free parameters, $f\sigma_8$ and $\beta$, in wavenumber ranges
$n\Delta k$ to $(n+1)\Delta k$ for $n=0,1,2,\dots$. The multi-tracer
approach with density and velocity improves the constraints on $\beta$
at large scales. The TAIPAN survey can measure growth rates, $\beta$
and $f\sigma_8$, to 20--30 per cent in each bin, and the WALLABY+WNSHS
surveys produce measurements with about 15 per cent precision in each
bin. One caveat is that the classical approximation in the Fisher
matrix could break down, giving inaccurate forecasts, at low $k$
comparable to the size of the surveys; we leave the work beyond classical
approximation to future studies. The two-field constraints on growth
rates predict large improvements for $k\le 0.1\,h\mathrm{Mpc}^{-1}$ by
evading the cosmic variance limit. These are the ranges where we can
recognise possible deviations from the $\Lambda$CDM model,
distinguishing them from non-linear dynamics, non-linear
redshift-space distortions, or scale-dependent galaxy bias.

\section{Summary and discussion}
\label{sec:summary}
We summarise our conclusions as follows:
\begin{itemize}
\item{We have improved the model for the auto- and cross-power spectra
  of galaxy density contrast and line-of-sight peculiar velocity
  fields. We show that the density-velocity cross-power spectrum and
  the velocity auto-power spectrum contain strong redshift-space
  distortions. We introduce a new damping term to the model equations,
  which needs to be considered in future velocity analyses to
  avoid biased results. }

\item{We compare the model equations for the power spectra with the
  GiggleZ simulation using the subhaloes. We calibrate the
  nearest particle method to compute reliable velocity power spectra
  (Appendix~\ref{sec:appendix_power_spectra}). The comparison shows
  that our model agrees well with the simulation for $k\la
  0.2\,h\mathrm{Mpc}^{-1}$.}

\item{We derive the Fisher matrix formula for multiple correlated fields
  when the noise terms vary with distance, including a pair of galaxy
  density and velocity fields whose shot noise and velocity
  measurement error increase with distance
  (Appendix~\ref{sec:fisher_matrix_derivation}). The derivation
  reminds us that the Fisher matrix uses the classical approximation
  which breaks down at low $k$. Since much of the constraint from
  peculiar velocity comes from low $k$, it is worthwhile to reexamine
  the validity of the classical approximation for peculiar velocities in
  the future. We also derive an equivalent Fisher matrix formula
  written in terms of the covariance of power spectra estimators
  (Appendix~\ref{sec:fisher_matrix_covp}). }

\item{When the number density of the peculiar velocity sample is the
  same as the redshift sample, $n_u=n_g$, the peculiar velocity survey
  improves the constraints on growth rates, $f\sigma_8$ and $\beta$ by
  more than a factor of 2 at $k_\mathrm{max}=0.2\,h\mathrm{Mpc}^{-1}$
  and about a factor of 5 for $k_\mathrm{max}=0.1\,h\mathrm{Mpc}^{-1}$
  for redshifts less than 0.1 (Section~\ref{sec:fisher_results}). Peculiar
  velocity surveys can also measure the galaxy-matter
  cross-correlation coefficient $r_g$ very precisely. With
  redshift-space distortions alone, in contrast, $f\sigma_8$ and $r_g$
  are highly degenerate, weakening the constraint by an order of
  magnitude if $r_g$ is a free parameter.}

\item{Lack of knowledge of the damping constants of redshift-space
  distortions, $\sigma_g$ and $\sigma_u$, degrade the constraint on
  $f\sigma_8$ by about 50 per cent, e.g., from 3 per cent to 4.5
    percent for the forecast for the WALLABY survey. Further
  development of the theory of velocity power spectrum in redshift space
  is necessary to extract accurate parameters from future peculiar
  velocity surveys. Uncertainties in the other cosmological parameters
  do not affect the constraints on the growth rates when the
  \textit{Planck} CMB data is added.}

\item{Future peculiar velocity surveys, TAIPAN, WALLABY and WNSHS,
  will constrain the growth rate $f\sigma_8$ with 3 per cent
  precision at low-redshift $z\le 0.05$. The growth rate can also be
  measured at different scales. In wavenumber bins with width $\Delta
  k = 0.01\,h\mathrm{Mpc}^{-1}$, $f \sigma_8$ and $\beta$ can be
  measured to $20-30$ per cent by the TAIPAN survey, and about $15$
  per cent by the WALLABY+WNSHS surveys in the range $0.01\,
  h\mathrm{Mpc}^{-1} \le k \le 0.1\,h\mathrm{Mpc}^{-1}$. These
  constraints on very large scales are largely improved, compared to
  redshift measurements alone, by the strength of peculiar velocity
  surveys that cosmic variance is not a fundamental limit. We can
    use the physical relation between the measured density and
    velocity to measure the growth rate $\beta$. These strong
  constraints contribute to constraining dark energy and modified
  gravity, which can have various growth rate functions of time and
  scale.}
\end{itemize}

We show that peculiar velocity surveys provide competitive growth rate
measurements at low redshift, $z \la 0.1$. Future peculiar velocity
surveys measure both redshifts and velocities with high number
densities. Their biggest strength is in measuring the growth rate as a
function of scale, which provides independent constraints on dark
energy and modified theories of gravity. These features of low
redshift and scale dependence are complimentary to large high redshift
surveys which measures growth rates as a function of time.

%
% Acknowledgments, References
%
\section*{Acknowledgements}
We thank Takahiko Matsubara, Wojciech Hellwing, Teppei Okumura, and
Maciej Bilicki for useful discussions. This research was conducted by
the Australian Research Council Centre of Excellence for All-sky
Astrophysics (CAASTRO), through project number CE110001020. We
acknowledge the support of the Australian Research Council through
Future Fellowship awards, FT110100639 (CB) and FT100100595 (TMD). GP
acknowledges support from S. Wyithe's ARC Laureate grant
(FL110100072).

\bibliographystyle{mn2e}
\bibliography{velocity}

%
% Appendix
%
\appendix

%
% A. Nearest particle method for velocity power spectrum
%
\section{Nearest particle method for velocity power spectrum}
\label{sec:appendix_power_spectra}
We calculate the auto- and cross-power spectra of subhalo density and
peculiar velocity by Fourier transforming the density and velocity fields
assigned on regular grid points. In this appendix, we explain how we
calculate the discrete fields and correct for the smoothing and
aliasing due to finite grid points.  The goal is to minimise the
numerical effect that depends on the grid resolution. We first review
the procedure for the density field by \citet{2005ApJ...620..559J} in
Section~\ref{sec:pgg_correction}, and then explain an analogous
procedure for the velocity field in
Section~\ref{sec:velocity_correction}.

Let us consider $N$ particles in a periodic box of length $L$ on a
side, at positions $\bmath{x}_p$, with line-of-sight velocity $u_p$
for $p=1,\dots,N$. We sample discrete densities and velocities on
$N_\mathrm{grid}^3$ regular grid points at $\bmath{x}_I$ for $I=1,
\dots, N_\mathrm{grid}^3$.

\begin{figure*}
\centering
\includegraphics[width=174mm]{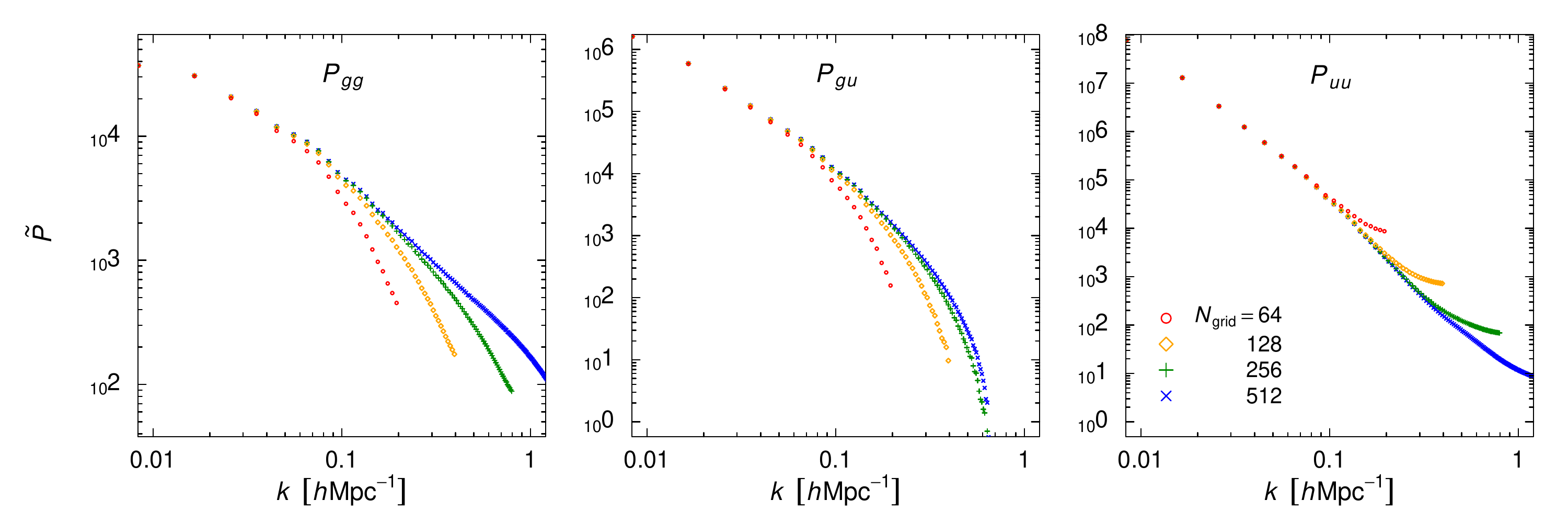}
\caption{Power spectra before corrections. The auto- and cross-power
  spectra of subhalo density and line-of-sight peculiar velocity are
  calculated on grids with different resolutions $N_\mathrm{grid} =
  64,128,256$ and $512$. \textit{Left} panel is the subhalo-subhalo
  auto-power spectrum $\tilde{P}_{gg}$, \textit{middle} is the
  subhalo-velocity cross power $\tilde{P}_{gu}$, and \textit{right} is
  the velocity-velocity auto-power spectrum $\tilde{P}_{uu}$. See the
  caption of Fig.~\ref{fig:power_spectra_real} for the units of the
  power spectra.}
\label{fig:p_before_correction}
\end{figure*}

\begin{figure*}
\centering
\includegraphics[width=174mm]{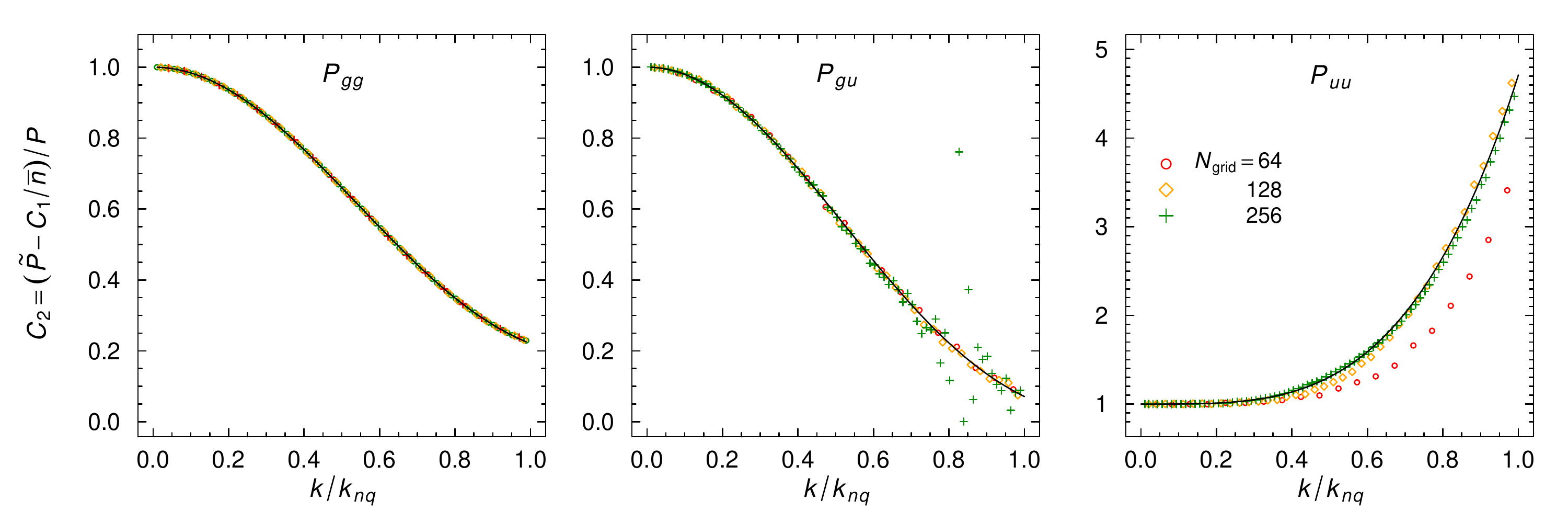}
\caption {The ratio of power spectrum calculated on a grid to the
  true power spectrum (the shot noise $C_1$ is first subtracted from
  the subhalo auto power). The black curves are the $C_2$ functions we
  use to correct the gridding effect (equations \ref{eq:c2gg},
  \ref{eq:c2gu}, and \ref{eq:c2uu}).  }
\label{fig:c2_function}
\end{figure*}

\begin{figure*}
\centering
\includegraphics[width=174mm]{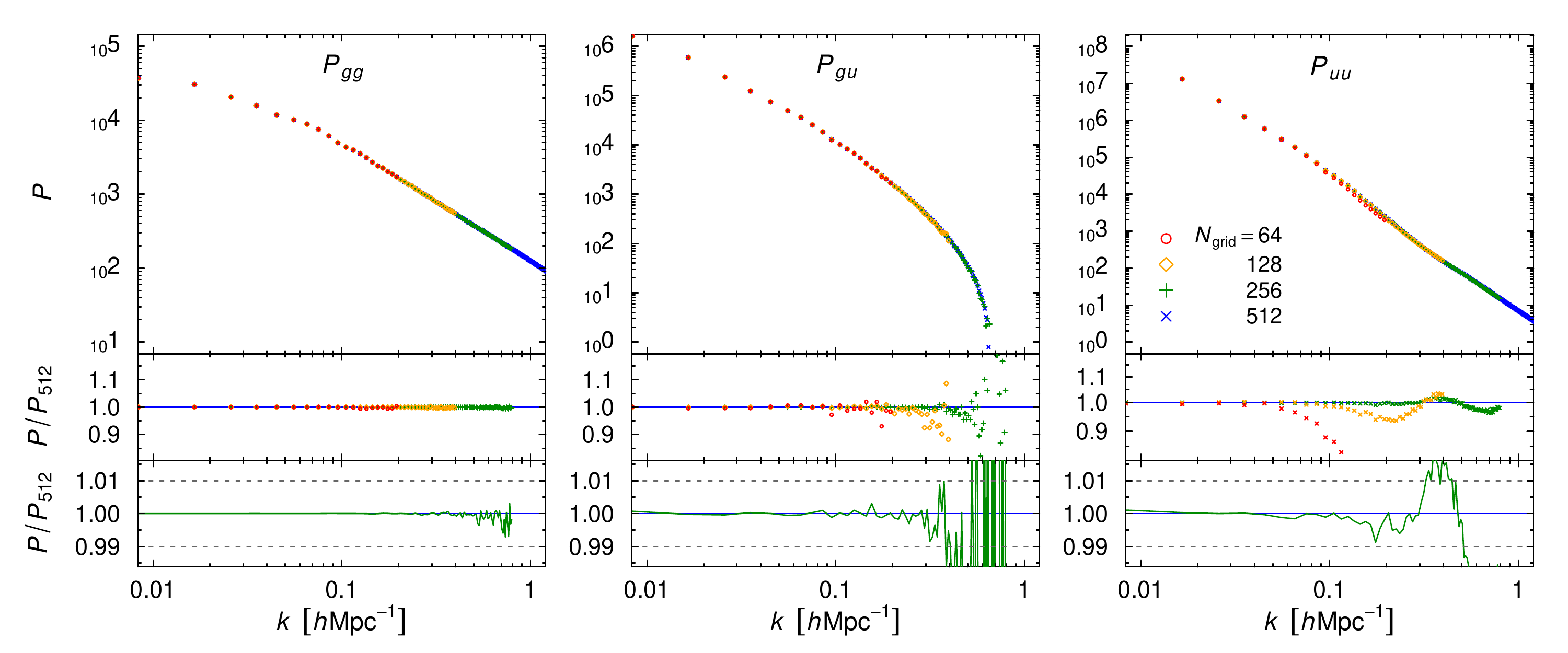}
\caption{Power spectra after corrections (\textit{top} panels), and
  relative difference compared with $N_\mathrm{grid} = 512$ power
  spectra at 10 per cent level (\textit{middle} row) and 1 per cent
  level for $N_\mathrm{grid}=256$. (\textit{bottom}
  row). $N_\mathrm{grid} = 256$ converge within $1$ per cent for $k
  \le 0.3$. }
\label{fig:p_after_correction}
\end{figure*}

\subsection{Density power spectrum}
\label{sec:pgg_correction}
We use the standard clouds-in-cell (CIC) method to calculate the
density field \citep{1981csup.book.....H}. The power spectrum
calculated from this discrete density grid is well understood. We
summarise the procedure by \citet{2005ApJ...620..559J}. The number
density field before sampling is a sum of Dirac delta functions
$\delta_D$,
\begin{equation}
  n(x) = \sum_{p=1}^{N} \delta_D(\bmath{x} - \bmath{x}_p).
\end{equation}
This is the density field independent of gridding. The density field
sampled on a grid point $\bmath{x}_I$ becomes,
\begin{equation}
  \tilde{n}(x_I) = \int \! d^3 y \, W_g(\bmath{x}_I-\bmath{y}) n(y),
\end{equation}
where the window function $W_g$ for CIC is,
\begin{equation}
  W_g(\bmath{r}) = \prod_{i=1}^3 (1 - |r_i|/\Delta x), 
\end{equation}
if $|r_i| \le \Delta x \equiv L/N_\mathrm{grid}$ for all $i=1,2,3$,
and zero otherwise. The galaxy (subhalo) auto-power spectrum
calculated from the grid $\tilde{P}_{gg}$ is related to the true power
spectrum $P_{gg}$ in the following way:
\begin{equation}
  \tilde{P}_{gg}(\bmath{k}) = 
    \sum_{\bmath{n}} |\hat{W}_g(\bmath{k}+k_P \bmath{n})|^2 
                              P_{gg}(\bmath{k} + k_P \bmath{n}) 
    + \bar{n}^{-1} C_1(\bmath{k}),
\end{equation}
where the sum is over 3-dimensional integer lattice $\bmath{n} = (n_1,
n_2, n_3)$, $\hat{W}_g$ is the Fourier transform of $W_g$,
\begin{equation}
  \hat{W}_g(\bmath{k}) = 
    \prod_{i=1}^3 \left[ \frac{\sin(\pi k_i/k_P)}{\pi k_i/k_P} \right]^2,
\end{equation}
where $k_P \equiv 2\pi N_\mathrm{grid}/L$ is the period of Fourier
modes, $\bar{n} \equiv N/L^3$ is the mean number density, and the
smoothed shot noise term is,
\begin{eqnarray}
  C_1(\bmath{k}) &\equiv& 
     \sum_{\bmath{n}} \left| \hat{W}_g(\bmath{k} + k_P \bmath{n}) \right|^2  \\
     &=& \prod_{i=1}^3 \left[ 1 - \frac{2}{3} \sin^2(\pi k_i/k_P) \right].
\end{eqnarray}
Discrete sampling introduces a periodicity in Fourier space with a
period of $k_P$, and all modes higher than the Nyquist frequency
$k_\mathrm{Nq} \equiv \pi N_\mathrm{grid}/L = k_P/2$ are added to the
modes in $|k_i| \le k_\mathrm{Nq}$. This is known as aliasing.
Because the power spectrum beyond the Nyquist frequency is not known a
priori, the power spectrum is extrapolated by a power law beyond
$k_\mathrm{Nq}$,
\begin{equation}
  P_{gg}(\bmath{k} + k_P \bmath{n} ) \approx 
    P_{gg}(k)(|k + k_P \bmath{n}|/k)^{n_\mathrm{eff}}.
\end{equation}
This $n_\mathrm{eff}$ can be determined iteratively, but we obtained sufficient
accuracy by setting $n_\mathrm{eff} = -1.6$. This extrapolation makes
it possible to calculate the smoothing factor $C_2$:
\begin{equation}
  \label{eq:Pgg_smoothing}
  \tilde{P}_{gg}(\bmath{k}) \approx
     C_{2gg}(\bmath{k}) P_{gg}(\bmath{k}) + \bar{n}^{-1} C_1(\bmath{k}),
\end{equation}
where,
\begin{equation}
  \label{eq:c2gg}
  C_{2gg}(\bmath{k}) \equiv 
    \sum_{\bmath{n}} |\hat{W}_g(\bmath{k} + k_P \bmath{n})|^2 \, 
    (|\bmath{k} + k_P \bmath{n}|/k_P)^{n_\mathrm{eff}}.
\end{equation}
The corrected power spectrum $P_{gg}(\bmath{k})$ is calculated from
the grid power spectrum $\tilde{P}_{gg}(\bmath{k})$ by first
subtracting the shot noise, $\bar{n}^{-1} C_1(\bmath{k})$, and then
divided by $C_{2gg}(\bmath{k})$. We spherically-averaged the power
spectrum after the correction.

In the left panel of Fig.~\ref{fig:p_before_correction}, we plot the
spherically-averaged subhalo power spectrum before corrections,
$\tilde{P}_{gg}$, for different grid resolutions
$N_\mathrm{grid}=64,128,256$, and $512$. The power spectra are plotted
up to the Nyquist frequencies $k_\mathrm{Nq} = 0.20 \,
(N_\mathrm{grid}/64)$ for $L=1h^{-1} \mathrm{Gpc}$. The grid
power spectrum is smoothed near $k_\mathrm{Nq}$. In the left panel of
Fig.~\ref{fig:c2_function}, we plot the ratio of grid power spectra
after shot noise subtraction to the true power spectra, which is the
$C_{2gg}$ function in equation~(\ref{eq:Pgg_smoothing}). We first
subtract the shot noise $C_1(\bmath{k})$ from
$\tilde{P}_{gg}(\bmath{k})$, take its spherical average, and then
divide that by the true spherically-averaged power spectrum
$P_{gg}(k)$, which is the corrected power spectrum for
$N_\mathrm{grid}=512$ here. The black curve on top of those
simulation data is the spherically-averaged theoretical curve
(equation~\ref{eq:c2gg}); we calculated the equation on a
3-dimensional grid with $N_\mathrm{grid} = 256$, with finite sums for
$n_i = -2,-1,0$ and $1$, and then took the spherical average. Because
the function is rapidly decreasing with $k$, most of the aliases are
negligible. Only the $n_i=-1$ aliases have comparable magnitude. This
curve matches perfectly with the data points computed from simulation
subhaloes. Finally, in the left panel of
Fig.~\ref{fig:p_after_correction}, we plot the power spectrum after
the correction. The power spectra with different grid resolution are
consistent within 1 per cent for all $k \le k_{Nq}$.

\subsection{Field of nearest particle velocity}
\label{sec:velocity_correction}
We use the nearest particle velocity to calculate the grid velocity
field. For each grid point $x_I$, we find the particle nearest to the
grid point, and set the grid velocity field $\tilde{u}(x_I)$ equal to
the line-of-sight velocity of that particle. We apply this methods
  for subhaloes whose number density is much lower than the number of
  grid points. Other methods could be suitable in the other limit of
  many particles per grid point; see also
  Section~\ref{sec:np_vpower_discussion}. We use the kd-tree
algorithm to find the nearest neighbour computationally efficient. The
smooth field before gridding, for this method, is the piecewise
constant velocity field which is equal to the particle velocity in
each Voronoi cell \citep{1996MNRAS.279..693B},
\begin{equation}
  u(\bmath{x}) = \sum_a u_a \chi_a(\bmath{x}),
\end{equation}
where $\chi_a(\bmath{x}) = 1$ if the particle $a$ is the nearest
particle of point $\bmath{x}$, and $\chi_a(\bmath{x}) = 0$
otherwise. The window function of this nearest particle
method is the Dirac delta function $W_u(\bmath{r}) =
\delta_D(\bmath{r})$:
\begin{equation}
  \label{eq:velocity_kernel}
  \tilde{u}(\bmath{x}_I) = \int \! d^3 y \,
    W_u(\bmath{x}_I - \bmath{y}) u(\bmath{y}).
\end{equation}
The Fourier transform of the window function is $\hat{W}_u(\bmath{k})
= 1$.

The advantages of our method are that (a) the velocity field on the
grid converges to a meaningful velocity field $u(\bmath{x})$ as
$N_\mathrm{grid} \rightarrow \infty$, and (b) the assignment to grid
points is done by a convolution
(equation~\ref{eq:velocity_kernel}). Since nearest neighbour exists at
any grid point, we do not have the problem of empty cells that the
velocity field becomes undefined. Another common difficulty is that
the smoothing kernel becomes spatially varying, not in the form of a
convolution. Typically, the normalisation factor depends on $x_I$,
which makes the kernel in equation~(\ref{eq:velocity_kernel}) a
function of $\bmath{x}_I$ as well, $W_u(\bmath{x}_I - \bmath{y},
\bmath{x}_I)$. This would make the correction complicated, because the
smoothed power $\tilde{P}(\bmath{k})$ at wavenumber $\bmath{k}$, would
then depend on the power at all wavenumbers $P(\bmath{k}')$,
not only on the power at the same wavenumber $P(\bmath{k})$.

In the middle and right panels of Fig.~\ref{fig:p_before_correction}, we
plot the subhalo density-velocity cross-power $\tilde{P}_{gu}$ and the
velocity-velocity auto-power $\tilde{P}_{uu}$, respectively,
calculated from CIC density and nearest particle velocity fields
before any corrections. The cross power has a smoothing, and the
velocity auto power has an increase in power due to aliasing.

\subsubsection{Cross-power correction}
In the middle panel of Fig.~\ref{fig:c2_function}, we plot the
angle-averaged cross-power calculated for $N_\mathrm{grid}=64,128,$
and $256$ divided by the angle-averaged $N_\mathrm{grid}=512$
cross-power after the correction we will describe here. (The
angle-average is only performed in the upper-half of $k$ space, i.e., 
$\int_0^1 d\mu$.)  Although we naively expect this ratio to be
$\hat{W}_g \hat{W}_u = \hat{W}_g$, because the density is smoothed by
a factor of $\hat{W}$ and the velocity field is smoothed by
$\hat{W}_u$, we find a smoothing closer to $\hat{W}_g^{2.5}$. We fit
the points by an empirical formula,
\begin{equation}
  \label{eq:c2gu}
  C_{2gu}(k) \equiv \left[ \frac{\sin (\pi k/k_P)}{\pi k/k_P} \right]^5 
                \left[ 1 - 0.27 (k/k_{Nq})^5 \right],
\end{equation}
where $k_P= 2 \pi N_\mathrm{grid}/L$ and $k_{Nq} = k_P/2$ are the same
as those in the previous section. This fitting formula is plotted by
a black curve in the figure. We use this function to correct the cross-power,
\begin{equation}
  P_{gu}(\bmath{k}) = \tilde{P}_{gu}(\bmath{k})/C_{2gu}(k)
\end{equation}
In the middle panel of Fig.~\ref{fig:p_after_correction}, we plot the
corrected cross-power. Although there are some scatter at high $k$,
the $N_\mathrm{grid}=256$ cross power converge within 1 per cent for
$k \le 0.3\,h\mathrm{Mpc}^{-1}$.

\subsubsection{Velocity auto-power correction}
In the right panel of Fig.~\ref{fig:c2_function}, we plot the ratio of
the angle-averaged velocity auto-power spectra calculated on a
grid to that calculated with $N_\mathrm{grid}=512$ with corrections.
We do not subtract a component analogous to the shot noise $C_1$. This
could be the reason that the $N_\mathrm{grid}=64$ points do not match
with other points.  The $C_2$ function analogous to
equation~(\ref{eq:c2gg}) for velocity-velocity power spectrum with
$W_u=1$ is,
\begin{eqnarray}
  \label{eq:c2uu_function_sum}
  C_{2uu} &=& \sum_{\bmath{n}} (|\bmath{k}+k_P \bmath{n}|/k_P)^{n_\mathrm{eff} - 2} 
             \nonumber\\
         &=& 1 + (k/k_{Nq})^{n_\mathrm{eff}-2} \sum_{\bmath{n} \ne \bmath{0}}
             \left(\frac{\bmath{k}}{2k} + \bmath{n} \right)^{n_\mathrm{eff}-2}.
\end{eqnarray}
The exponent is $n_\mathrm{eff}-2$, because velocity power spectra has
an extra $k^{-2}$ factor compared to matter power spectra in linear
theory. This sum over integers converges very slowly, which makes it
impractical to calculate $C_{2uu}(\bmath{k})$ for all $\bmath{k}$. We
approximate the sum by a constant by fitting the points from the
simulation for $N=128$ and $256$,
\begin{equation}
  \label{eq:c2uu}
  C_{2uu} \approx 1 + 3.7 (k/k_{Nq})^{n_\mathrm{neff} - 2}.
\end{equation}
The constant prefactor is consistent with
equation~(\ref{eq:c2uu_function_sum}) for $\bmath{k}$ near the Nyquist
frequency. We correct the power spectrum with this formula,
\begin{equation}
  P_{uu}(\bmath{k}) = \tilde{P}_{uu}(\bmath{k})/C_{2uu}(k).
\end{equation}
The corrected power spectra is plotted in the right panel of
Fig.~\ref{fig:p_after_correction}. The velocity auto power with
$N_\mathrm{grid}$ also converges within 1 per cent for $k \le
0.3\,h\mathrm{Mpc}^{-1}$.

\subsection{Discussion on computing the velocity power spectrum}
\label{sec:np_vpower_discussion}
We presented our relatively simple method of using nearest particle
velocity to calculate velocity power
spectrum. \citet{2013arXiv1308.0886Z} independently used a similar
method to calculate the velocity power spectrum; see also their paper for
various numerical convergence tests. A drawback of this method is that
all high $k$ aliases add to low $k$ modes without smoothing, because
the sampling function in Fourier space is a constant, not a rapidly
declining function of $k$.

Calculating velocity power spectrum has technical difficulties that
do not exist for density power spectrum. Calculating velocity by
first assigning momentum on grids and then dividing them by density
using a fixed kernel (e.g., CIC) has two problems. One is that the
velocity becomes undefined if the density is zero. In the limit of
infinite grids, velocity becomes undefined almost everywhere, which
means there is no proper convergence as we increases the number of
grids. Using adaptive kernel, in which the kernel length increases at
low density regions, can avoid the problem of undefined velocity
\citep[e.g.,][]{2012MNRAS.422..926M}, but the resulting velocity field
is smoothed in a complicated way, for which it is difficult (not
necessarily impossible, but at least computationally expensive) to
deconvolve the kernel smoothing. The other problem is that, since the
Fourier transformation is a volume integral, mixing mass-weighted
average within grid cells make the convergence, as the number of grids
increase, inefficient. \citet{2009PhRvD..80d3504P} use volume-weighted
average by calculating the volume from local Delaunay tessellation
using volumes that are entirely inside a grid cell. This improves the
accuracy and convergence, but Delaunay tessellations spanning several
cells are not treated accurately for simplicity, which is sufficient
for a large number of particles, but becomes problematic for sparse
samples, such as galaxies or haloes. Integration of velocity field to
a grid cell is necessary \citep{1996MNRAS.279..693B}.

The Delaunay Tessellation Field Estimator (DTFE)
software\footnote{\texttt{http://www.astro.rug.nl/\~{}voronoi/DTFE/dtfe.html}}
\citep{2000A&A...363L..29S, 2011ascl.soft05003C} is a publicly
available code that calculates velocity field interpolated by the
Delaunay tessellation, and integrating the field numerically on grids
with the Monte Carlo approach. The DTFE software works well for
velocity power spectrum, too \citep{2012MNRAS.427L..25J}.  This Monte Carlo
integration is a reasonable method to suppress the high $k$ aliases,
which can be added to improve our nearest particle method in the
future if necessary.

%
% B. Theorems of Fisher Matrix
%
\section{The Fisher matrix for multiple tracers}
\label{sec:fisher_matrix_appendix}
In this appendix, we first derive the Fisher matrix with the classical
approximation in Section~\ref{sec:fisher_matrix_derivation}, and then
drive other forms of the Fisher matrix in
Sections~\ref{sec:fisher_matrix_transformation}--\ref{sec:fisher_matrix_covp}.

\subsection{Derivation}
\label{sec:fisher_matrix_derivation}
We summarise the derivation of Fisher matrix using the `classical
approximation' \citep{1997MNRAS.289..285H, 2012MNRAS.420.2042A}. This
approximation simplifies the Fisher Matrix with spatially
inhomogeneous noise to a form similar to that with constant noise.

Let $\phi_a(\bmath{x})$ ($i=1, \dots, N$) be $N$ real Gaussian fields
in configuration space which have zero mean, $\langle \phi_a(\bmath{x})
\rangle = 0$. They can be a pair of galaxy density field and
line-of-sight peculiar velocity, $N$ galaxy density fields with
different biases, or any multiple tracers of a random Gaussian
field. We apply equation~(\ref{eq:fisher_gaussian}) to this
continuously infinite number of Gaussian variables labelled by the
position $\bmath{x}$. The mean vector is zero, $\bmath{\mu} =
\bmath{0}$, and the covariance matrix $\mathbfss{C}$ is labelled by two
positions $C_{mn} \rightarrow \Sigma_{ab}(\bmath{x}, \bmath{y}) \equiv
\langle \phi_a(\bmath{x}) \phi_b(\bmath{y})\rangle$, which are the
auto- or cross-two-point correlation functions. In this continuous
limit, sums over matrix indexes are replaced by integrals;
equation~(\ref{eq:fisher_gaussian}) becomes,
\begin{eqnarray}
  \label{eq:fisher_matrix_contineous}
  F_{ij}\!\!\!\!&=&\!\!\!\! \frac{1}{2} \int\! d^3x \,d^3x' \, d^3y \, d^3y' \nonumber \\ 
        & &\!\!\!\! \mathrm{tr} \left[ 
    \Sigma^{-1}(\bmath{x}, \bmath{x}')
    \frac{\partial \Sigma(\bmath{x}', \bmath{y})}{\partial \theta_i}
    \Sigma^{-1}(\bmath{y}, \bmath{y}')
    \frac{\partial \Sigma(\bmath{y}', \bmath{x})}{\partial \theta_j}
    \right].
\end{eqnarray}
The inverse function is defined as,
\begin{equation}
  \label{eq:sigma_inv_def1}
  \int\! d^3y\, \Sigma^{-1}(\bmath{x}, \bmath{y}) \Sigma(\bmath{y}, \bmath{z})
    = I_N \delta_D(\bmath{x}-\bmath{z}),
\end{equation}
and
\begin{equation}
  \label{eq:sigma_inv_def2}
  \int\! d^3y\, \Sigma(\bmath{x},\bmath{y}) \Sigma^{-1}(\bmath{y},\bmath{z}) 
    = I_N \delta_D(\bmath{x}-\bmath{z}),
\end{equation}
where $I_N$ is the $N \times N$ unit matrix.  

The covariance function $\Sigma(\bmath{x}, \bmath{y})$ contains 
translationally invariant correlation functions $\xi_{ab}$, and spatially
uncorrelated noise terms $N_{ab}$,
\begin{equation}
  \label{eq:sigma_ab}
  \Sigma_{ab}(\bmath{x}, \bmath{y}) =
\zeta_{ab}(\bmath{x}-\bmath{y}) + N_{ab}(\bmath{x})
\delta_D(\bmath{x}-\bmath{y}),
\end{equation}
where $\delta_D$ is the Dirac delta function.

For a pair of galaxy density contrast field, $\phi_1 = \delta_g$, and
line-of-sight velocity $\phi_2 = u$, the matrix of correlation functions is,
\begin{equation}
  \label{eq:sigma_matrix}
  \Sigma(\bmath{x}, \bmath{y}) = \left(
    \begin{array}{cc}
      \xi_{gg}(\bmath{x} - \bmath{y}) + N_g &
      \xi_{gu}(\bmath{x} - \bmath{y})  \\
      \xi_{ug}(\bmath{x} - \bmath{y})  &
      \xi_{uu}(\bmath{x} - \bmath{y}) + N_u
    \end{array}
  \right).
\end{equation}
The noise term for the density contrast, $N_g$, is the shot noise,
\begin{equation}
  N_{g}(\bmath{x}, \bmath{y}) = 
    n_g^{-1}(\bmath{x}) \delta_D(\bmath{x}-\bmath{y}),
\end{equation}
where $n_g(\bmath{x})$ is the smooth ensemble mean number density of
galaxies. Similarly, $N_g$ is the noise in peculiar velocity
measurement,
\begin{equation}
  N_{u}(\bmath{x}, \bmath{y}) = 
    n_u^{-1}(\bmath{x}) \sigma_\mathrm{u-noise}^2(\bmath{x}) 
      \delta_D(\bmath{x}-\bmath{y}),
\end{equation}
where $n_u(\bmath{x})$ is the mean number density of galaxies with peculiar
velocity measurements, and $\sigma_\mathrm{vobs}$ is the observational error
in peculiar velocity per galaxy.

Using the fact that a two-point correlation function $\xi_{ab}$ is the
Fourier transform of the corresponding power spectrum $P_{ab}$, we can
write $\Sigma(\bmath{x}, \bmath{y})$ with power spectra,
\begin{equation}
  \label{eq:sigma_k_x}
  \Sigma(\bmath{x}, \bmath{y}) = 
   \int\! \frac{d^3k}{(2\pi)^3} \tilde{\Sigma}(\bmath{k}, \bmath{x})
    e^{i\bmath{k}\cdot (\bmath{x}-\bmath{y})},
\end{equation}
where,
\begin{equation}
  \tilde{\Sigma}(\bmath{k}, \bmath{x})_{ab} \equiv
    P_{ab}(\bmath{k}) + N_{ab}(\bmath{x}).
\end{equation}
Equation~(\ref{eq:sigma_matrix}), for a case of density and velocity,
transforms to,
\begin{equation}
  \tilde{\Sigma}(\bmath{k}, \bmath{x}) = \left(
    \begin{array}{cc}
      P_{gg}(\bmath{k}) + n_g^{-1}(\bmath{x}) & P_{gu}(\bmath{k}) \\
      P_{ug}(\bmath{k}) & P_{uu}(\bmath{k}) + n_u^{-1} \sigma_\mathrm{u-noise}^2
    \end{array}
\right). 
\end{equation}
[$P_{ug}(\bmath{k}) = \langle u(\bmath{k}) \delta_g(\bmath{k})^* \rangle$
is the complex conjugate of $P_{gu}(\bmath{k})$.]  For $N$ biased
galaxy tracers with biases $b_i$ and mean number density
$n_i(\bmath{x})$, $\tilde{\Sigma}$ matrix is, e.g.,
\begin{equation}
  \tilde{\Sigma}_{ab}(\bmath{k}, \bmath{x}) = 
    b_i b_j (1+\beta_i\mu^2)(1+\beta_j\mu^2) P_m(\bmath{k}) + 
    n_i^{-1}(\bmath{x})\delta_{ab},
\end{equation}
where $\beta_i \equiv f/b_i$, for the simplest case of the linear theory.

The delta functions in the noise terms enable us to replace
$\bmath{x}$ by $\bmath{y}$, if necessary,
\begin{equation}
  \label{eq:sigma_k_y}
  \Sigma(\bmath{x}, \bmath{y}) = 
   \int\! \frac{d^3k}{(2\pi)^3} \tilde{\Sigma}(\bmath{k}, \bmath{y})
    e^{i\bmath{k}\cdot (\bmath{x}-\bmath{y})}.
\end{equation}
A symmetry in correlation function, $\Sigma_{ab}(\bmath{x},\bmath{y})
= \Sigma_{ba}(\bmath{y}, \bmath{x})$, by definition, propagates to a
property that $\tilde{\Sigma}$ is a Hermitian matrix:
$\tilde{\Sigma}_{ab}(\bmath{k}, \bmath{x}) =
\tilde{\Sigma}_{ba}(\bmath{k}, \bmath{x})^*$.  The matrix
$\tilde{\Sigma}$ is a real symmetric matrix for density contrasts of
multi-tracers, but for density and peculiar velocity, the off-diagonal
term $P_{gu}$ is pure imaginary as a result of parity invariance, as
we discussed in Section~\ref{sec:velocity_powerspectrum}.

The `classical approximation' \citep{1997MNRAS.289..285H} allows us to
approximate the inverse function by the inverse matrix in Fourier
space,
\begin{equation}
  \label{eq:sigma_inv_classical}
  \Sigma^{-1}(\bmath{x},\bmath{y}) \approx
   \int\! \frac{d^3k}{(2\pi)^3}
   \tilde{\Sigma}(\bmath{k}, \bmath{x})^{-1} e^{i\bmath{k}\cdot(\bmath{x}-\bmath{y})},
\end{equation}
where $\tilde{\Sigma}(\bmath{k}, \bmath{x})^{-1}$ is the inverse
matrix of $\tilde{\Sigma}(\bmath{k}, \bmath{x})$ for fixed $\bmath{x}$
and $\bmath{k}$.  To check that is approximately an inverse function,
equations (\ref{eq:sigma_k_y}, \ref{eq:sigma_inv_classical})
substituted to the left-hand side of equation
(\ref{eq:sigma_inv_def1}) gives,
\begin{equation}
 \label{eq:sigma_inv_sigma}
  \int\! \frac{d^3 k}{(2\pi)^3} \tilde{\Sigma}(\bmath{k}, \bmath{x})^{-1}
    \tilde{\Sigma}(\bmath{k}, \bmath{z}) e^{i\bmath{k}\cdot(\bmath{x}-\bmath{z})}.
\end{equation}
If (a) $\bmath{x}\approx\bmath{z}$, such that the noise terms
$N(\bmath{x})$ and $N(\bmath{z})$ are approximately equal to each
other, then $\tilde{\Sigma}(\bmath{k}, \bmath{x})^{-1}
\tilde{\Sigma}(\bmath{k}, \bmath{z}) \approx I_N$ makes equation
(\ref{eq:sigma_inv_sigma}) equal the right-hand side of equation
(\ref{eq:sigma_inv_def1}). If (b) $\bmath{x}$ and $\bmath{z}$ are far
enough from each other, then
$e^{i\bmath{k}\cdot(\bmath{x}-\bmath{z})}$ is a rapidly oscillating
function of $\bmath{k}$, which makes the integral in
equation~(\ref{eq:sigma_inv_sigma}) approximately zero --- again the
equation (\ref{eq:sigma_inv_def1}) is satisfied. Because the integrand
depends on $\bmath{k}$ through $P_{ab}(\bmath{k})$, the approximation
that the rapid oscillation makes the integral vanishing is reasonable
if $\int\! \frac{d^3k}{(2\pi)^3} P_{ab}(\bmath{k})
e^{i\bmath{k}\cdot(\bmath{x}-\bmath{y})} =
\xi_{ab}(\bmath{x}-\bmath{z}) \approx 0$. In summary, the
approximation used here, called the classical approximation,
is valid if either (a) or (b) is satisfied for all values of
$\bmath{x}-\bmath{y}$. This means that the noise terms are
approximately constant within the coherence length, where the
two-point correlation functions are not
negligible. Equation~({\ref{eq:sigma_inv_def2}}) can be shown in the
same way. Although this approximation is usually satisfied for
redshift surveys, it requires more caution for peculiar velocity
surveys due to larger coherence length, and distant dependent
observational noise term $\sigma_\mathrm{vobs}$.

Substituting the inverse function
(equation~\ref{eq:sigma_inv_classical}) into the Fisher matrix 
equation~(\ref{eq:fisher_matrix_contineous}) gives,
\begin{eqnarray}
  F_{ij} &=& \frac{1}{2} 
             \int\! d^3 x\, d^3 y \frac{d^3 k}{(2\pi)^3} \frac{d^3 q}{(2\pi)^3} 
              e^{i(\bmath{k}-\bmath{q})\cdot(\bmath{x}-\bmath{y})} \nonumber \\
        & & \mathrm{tr} \left[
            \tilde{\Sigma}^{-1}(\bmath{k}, \bmath{x}) 
            \frac{\partial \tilde{\Sigma}}{\partial \theta_i}(\bmath{k})
            \tilde{\Sigma}^{-1}(\bmath{q}, \bmath{y}) 
            \frac{\partial \tilde{\Sigma}}{\partial \theta_j}(\bmath{q})
            \right],
\end{eqnarray}
where we use the assumption that the noise terms are determined from
observations, not directly related to cosmological parameters, i.e.,
\begin{equation}
  \label{eq:derivative-condition}
  \frac{\partial \tilde{\Sigma}(\bmath{k}, \bmath{x})}{\partial \theta_i} =
    \frac{\partial P(\bmath{k})}{\partial \theta_i},
\end{equation}
is a function of $\bmath{k}$ only. Using the same argument of the
classical approximation, the $k$ and $q$ integrals are negligible for
large $\bmath{x}-\bmath{y}$ due to rapid oscillation of the
exponential term. Therefore, the dominant contribution comes from
$\bmath{x}\approx\bmath{y}$, which allows us to replace,
$\tilde{\Sigma}(\bmath{q}, \bmath{y})^{-1} \approx
\tilde{\Sigma}(\bmath{q}, \bmath{x})^{-1}$. We can then rearrange
the integral $\int d^3x d^3y = \int d^3 x d^3 (x-y)$ and perform the
$d^3(x-y)$ integral. This gives our final result for the Fisher
Matrix,
\begin{equation}
 \label{eq:fisher_matrix_tr}
  F_{ij} = \frac{1}{2} \int \! \frac{d^3 x \, d^3 k}{(2\pi)^3} \,
    \mathrm{tr} \left[ 
      \tilde{\Sigma}(\bmath{k}, \bmath{x})^{-1}
      \frac{\partial \tilde{\Sigma}}{\partial \theta_i}
      \tilde{\Sigma}(\bmath{k}, \bmath{x})^{-1}
      \frac{\partial \tilde{\Sigma}}{\partial \theta_j}
    \right],
\end{equation}
where all $\tilde{\Sigma}$ are evaluated at $\bmath{k}$ and
$\bmath{x}$.  For a single field, this reduces to the Fisher matrix by
\citet{1997PhRvL..79.3806T} with \citet*{1994ApJ...426...23F} minimum
variance. The virtue of this derivation, starting from the Gaussian
fields in configuration space, is that this minimum variance appears
automatically, and the generalisation to multiple fields is straight
forward.

\subsection{Isomorphic transformation}
\label{sec:fisher_matrix_transformation}
The Fisher matrix is invariant under any invertible linear transformation
(isomorphism) $\phi' = A \phi$ between statistically translational
invariant fields if that transformation does not include any uncertain
parameters $\theta_i$. For example, \citetalias{2004MNRAS.347..255B}
uses line-of-sight velocity gradient $\phi' \equiv (\delta_g, \partial
u / \partial r)$, instead of velocity, $\phi=(\delta_g, u)$; we show
that the Fisher matrix for $\phi'$ is exactly equal to that for $\phi$.
This is one of the two steps that their Fisher matrix is exactly
equal to what we use in this paper.

Since our formalism is based on an assumption that the field is
translationally invariant (e.g., equation~\ref{eq:sigma_ab}), we require
that the linear transformation conserves translational
invariance. Such transformation is a convolution in
configuration space, which is a multiplication in Fourier space:
\begin{equation}
  \label{eq:linear_phi_transform}
  \phi'(\bmath{k}) = \mathbfss{A}(\bmath{k}) \phi(\bmath{k}).
\end{equation} 
We require that $\mathbfss{A}$ is a $N \times N$ matrix that has a
inverse $\mathbfss{A}^{-1}$, and it does not depend on parameters
$\theta_i$. The matrix of power spectra $\tilde{\Sigma}'$
(equation~\ref{eq:sigma_k_x}) for $\phi'$ is related to the original
matrix by,
\begin{equation}
  \tilde{\Sigma}'(\bmath{k}, \bmath{x}) = \mathbfss{A}(\bmath{k}) 
    \tilde{\Sigma}(\bmath{k}, \bmath{x}) \mathbfss{A}^\dagger(\bmath{k})
\end{equation}
where $\mathbfss{A}^\dagger$ is the Hermitian conjugate of
$\mathbfss{A}$. Such transformation does not change the trace:
\begin{equation}
  \mathrm{tr}\left[ 
    \Sigma'^{-1} \frac{\partial \Sigma'}{\partial \theta_i}
    \Sigma'^{-1} \frac{\partial \Sigma'}{\partial \theta_j}
  \right] =
  \mathrm{tr}\left[ 
    \Sigma^{-1} \frac{\partial \Sigma}{\partial \theta_i}
    \Sigma^{-1} \frac{\partial \Sigma}{\partial \theta_j}
  \right],
\end{equation}
for each value of $\bmath{k}$ and $\bmath{x}$ because all
$\mathbfss{A}$ matrices cancels their inverse matrices in the
trace. This proves what we stated at the beginning of this section:

\subsubsection{Lemma} The Fisher matrix (equation~\ref{eq:fisher_matrix_tr}) is
invariant under any isomorphism, $\phi' = \mathbfss{A} \phi$
(equation~\ref{eq:linear_phi_transform}), if it does not contain any of
the parameters $\theta_i$. The Fisher matrix formula with $\phi'$ is exactly
equal to that with $\phi$.

\subsubsection{Examples of isomorphism}
\label{sec:examples_isomorphism}
The aforementioned example of using velocity gradient
\citepalias{2004MNRAS.347..255B} is a transformation:
\begin{equation}
  \phi'(\bmath{k}) = \mathbfss{A} \phi = 
   \left(\begin{array}{cc}
      1 &  0\\ 
      0 & \mathrm{i}k\mu
   \end{array} \right)
   \left(\begin{array}{c}
      \delta_g(\bmath{k}) \\
      u(\bmath{k})
   \end{array}\right),
\end{equation}
with flat-sky approximation, where $\mu$ is the cosine of the angle
between $\bmath{k}$ and the fixed line-of-sight direction. [This
  matrix is invertible for $\bmath{k} \ne 0$, and
  $\bmath{k}=\bmath{0}$ mode of velocity is zero, not carrying any
  cosmological information.] The covariance matrix transforms as,
\begin{equation}
  \label{eq:sigma_u_prime}
  \tilde{\Sigma}' = \mathbfss{A}\tilde{\Sigma}\mathbfss{A}^\dagger
   = \left(\begin{array}{cc}
      \tilde{\Sigma}_{11} & -\mathrm{i}k\mu \tilde{\Sigma}_{11}  \\
      \mathrm{i}k\mu \tilde{\Sigma}_{22} & k^2 \mu^2 \tilde{\Sigma}_{22}
   \end{array} \right).
\end{equation}
Although the covariance matrices $\tilde{\Sigma}$ and
$\tilde{\Sigma}'$ look different, the Fisher matrices calculated from
those are exactly equal to each other.

We can also remove the imaginary number $\mathrm{i}$ from cross-power
spectra without changing the Fisher matrix. Applying a matrix,
\begin{equation}
  \label{eq:u_tilde}
  \tilde{\phi}(\bmath{k}) \equiv 
    \left(\begin{array}{c}
      \delta_g(\bmath{k}) \\
      \tilde{u}(\bmath{k})
   \end{array}\right)
  \equiv \mathbfss{A} \phi = 
   \left(\begin{array}{cc}
      1 & 0 \\
      0 & \mathrm{i}
   \end{array} \right)
   \left(\begin{array}{c}
      \delta_g(\bmath{k}) \\
      u(\bmath{k})
   \end{array}\right),
\end{equation}
transforms the cross power to a real function. Our lemma guarantees
that the Fisher matrix remains exactly the same.

\subsection{Fisher matrix with power spectra covariance matrix}
\label{sec:fisher_matrix_covp}
We derive an equivalent form of the Fisher matrix that is written with a
covariance matrix of power spectra. Such Fisher matrices appear in
\citetalias{2004MNRAS.347..255B} and
\citet*{2009MNRAS.397.1348W}. \citet{2009MNRAS.397.1348W} checked that
two forms of Fisher matrix give the same numerical result. Here we
show that the two formulae are algebraically equivalent.

Let us introduce $N$ complex Gaussian variables with zero mean,
$\delta_a(\bmath{k}, \bmath{x})$, for each pair of $\bmath{k}$ and
$\bmath{x}$ that is uniquely characterised by the covariance matrix,
\begin{equation}
  \label{eq:cov_delta}
  \mathrm{Cov}(\delta_a, \delta_b) \equiv \langle \delta_a \delta_b^* \rangle
    = \tilde{\Sigma}_{ab}(\bmath{k}, \bmath{x}),
\end{equation}
\begin{equation}
  \label{eq:cov_delta_star}
  \mathrm{Cov}(\delta_a, \delta_b^*) = \langle \delta_a \delta_b \rangle = 0
\end{equation}
 where $\tilde{\Sigma}(\bmath{k}, \bmath{x})$ is the Hermitian matrix
 defined in equation~(\ref{eq:sigma_k_x}). These $\delta$ variables
 can be regarded as $\phi_a(\bmath{k})$ in Fourier space when the
 noise terms are spatially homogeneous. When the noise terms were
 position dependent, it much clearer to define $\delta_a$ as a pure
 mathematical tool that assists the derivation, than to make this
 exact derivation inexact by going though the classical
 approximation again. We have shown that cross-power spectra are
 either real or pure imaginary, and when they are pure imaginary they
 can be converted to real cross power without changing the Fisher
 matrix. We can therefore assume, without loss of generality, that
 $\tilde{\Sigma}$ is a real symmetric matrix.

The trace term in the Fisher Matrix
(equation~\ref{eq:fisher_matrix_tr}), with equation
(\ref{eq:derivative-condition}), can be rewritten as,
\begin{equation}
  \label{eq:rewriting_tr}
  \mathrm{tr}\left[ 
    \tilde{\Sigma}^{-1} \frac{\partial P}{\partial \theta_i} 
    \tilde{\Sigma}^{-1} \frac{\partial P}{\partial \theta_j} 
  \right] = 
  \frac{\partial P^*_{ab}}{\partial \theta_i}
    \Omega_{abcd} \frac{\partial P_{cd}}{\partial \theta_j},
\end{equation}
where,
\begin{equation}
  \Omega_{abcd} \equiv \tilde{\Sigma}^{-1}_{ac} \tilde{\Sigma}^{-1}_{db}.
\end{equation}
All repeated indices are summed over $1,\dots, N$. Consider
$\bmath{P}$ as a $N^2$ dimensional vector whose elements are $P_{ab}$,
and $\Omega_{abcd}$ as $N^2 \times N^2$ matrix;
equation~(\ref{eq:rewriting_tr}) is $(\partial
\bmath{P}^\dagger/\partial \theta_i) \Omega (\partial
\bmath{P}/\partial \theta_j)$ with this notation. The inverse matrix of 
$\Omega$ 
is the $\Xi$ matrix, defined as,
\begin{equation}
  \Xi_{abcd} \equiv \tilde{\Sigma}_{ac}\tilde{\Sigma}_{db}.
\end{equation}
This can be checked by a calculation $\Omega_{abcd} \Xi_{cdef} =
\delta_{ae}\delta_{bf}$ ($\delta_{ab}$ is the Kronecker's delta),
which means $\Omega \Xi = I_{N^2}$ in matrix notation. This $\Xi$
matrix turns out to be a covariance matrix of $\hat{P}_{ab}$ random
variables,
\begin{equation}
  \hat{P}_{ab} \equiv \delta_a \delta_b^*,
\end{equation}
because,
\begin{eqnarray}
  \mathrm{Cov}(\hat{P}_{ab}, \hat{P}_{cd}) &\equiv&
   \langle \delta_a \delta_b^* (\delta_c \delta_d^*)^* \rangle -
   \langle \delta_a \delta_b^* \rangle \langle (\delta_c \delta_d^*)^* \rangle
     \nonumber \\
      &=& 
   \langle \delta_a \delta_c^* \rangle \langle \delta_d \delta_b^* \rangle +
   \langle \delta_a \delta_d \rangle \langle \delta_b^* \delta_d^* \rangle
     \nonumber \\
   &=& \tilde{\Sigma}_{ac} \tilde{\Sigma}_{db} = \Xi_{abcd},
\end{eqnarray}
or $\mathrm{Cov}(\hat{\bmath{P}}, \hat{\bmath{P}}) = \Xi$ in matrix
notation, where the Isserlis' theorem or the Wick's theorem for
Gaussian variables is used.  We therefore got the other expression
equivalent to the equation~(\ref{eq:fisher_matrix_tr}),
\begin{equation}
  \label{eq:fisher_matrix_p_n2}
  F_{ij} = \frac{1}{2} \int\! \frac{d^3x d^3k}{(2\pi)^3} \,
  \frac{\partial \bmath{P}^\dagger}{\partial \theta_i} 
  \mathrm{Cov}(\hat{\bmath{P}}, \hat{\bmath{P}})^{-1}
  \frac{\partial \bmath{P}}{\partial \theta_j}.
\end{equation}
This is a multiplication of $N^2$-dimensional vectors of power spectra
with the covariance matrix of their $N^4$ pairs.

We now reduce the $N^2$-dimensional vector to $N(N+1)/2$-dimensional
vector using the assumption that $\tilde{\Sigma}_{ab}$ is a real symmetric
matrix. Let $\mathbfss{R}$ be a $N^2 \times N^2$ invertible matrix
with $1$ and $\pm 1/2$, such that
   $\bmath{P}' \equiv (\bmath{P}_\mathrm{sym}, \bmath{P}_\mathrm{asym}) = 
     \mathbfss{R} \bmath{P}$
become the symmetric and asymmetric combinations of $P$,
\begin{equation}
  P^\mathrm{sym}_{ab}  = \frac{1}{2}(P_{ab} + P_{ba}) 
     \quad\mbox{  for $a \le b$},
\end{equation}
\begin{equation}
  P^\mathrm{asym}_{ab} = \frac{1}{2}( P_{ab} - P_{ba} )  
     \quad\mbox{  for $a > b$}.
\end{equation}
The $N(N+1)/2$ dimensional vector $\bmath{P}_\mathrm{sym}$ contains
$N$ auto-power spectra and $N(N-1)/2$ symmetrized cross-power spectra.
$\bmath{P}_\mathrm{asym}$ contains $N(N-1)/2$ anti-symmetrized
cross-power spectra, which vanish by our assumption of symmetric power
spectra: $\bmath{P}_\mathrm{asym} = 0$. The vector of rearranged power
spectra estimator,
  $(\hat{\bmath{P}}_\mathrm{sym}, \hat{\bmath{P}}_\mathrm{asym}) 
    \equiv \hat{\bmath{P}}' 
    \equiv R \hat{\bmath{P}}$,
has a covariance matrix,
\begin{equation}
  \label{eq:p_cov}
  \Xi' \equiv \mathrm{Cov}(\hat{\bmath{P}}', \hat{\bmath{P}}')
       = R \Xi R^\dagger.
\end{equation}
The integrand of the Fisher matrix
equation~(\ref{eq:fisher_matrix_p_n2}) maintains the same form under
this rearrangement,
\begin{equation}
  \label{eq:xi_prime_n2}
  \frac{\partial \bmath{P}'^\dagger}{\partial \theta_i} \Xi'^{-1} 
  \frac{\partial \bmath{P}'}{\partial \theta_j}
  = 
  \frac{\partial \bmath{P}^\dagger}{\partial \theta_i} \Xi^{-1} 
  \frac{\partial \bmath{P}}{\partial \theta_j}.
\end{equation}
Finally, this $\Xi'$ matrix is block diagonal 
\begin{equation}
  \Xi' = \left(\begin{array}{cc}
           \Xi'_\mathrm{sym} & 0  \\
             0 & \Xi'_\mathrm{asym}
         \end{array} \right),
\end{equation}
where 
  $\Xi_\mathrm{sym}' \equiv 
     \mathrm{Cov}(\hat{\bmath{P}}_\mathrm{sym}, \hat{\bmath{P}}_\mathrm{sym})$,
and
  $\Xi_\mathrm{asym}' \equiv 
     \mathrm{Cov}(\hat{\bmath{P}}_\mathrm{asym}, \hat{\bmath{P}}_\mathrm{asym})$,
respectively, because the off-diagonal block vanishes,
\begin{equation}
  \mathrm{Cov}(\delta_a \delta_b^* + \delta_b \delta_a^*,
               \delta_c \delta_d^* + \delta_d \delta_c^*) = 0,
\end{equation}
by straight forward calculation using the Wick theorem and the
assumption $\tilde{\Sigma}_{ab} = \tilde{\Sigma}_{ba}$.  Therefore the
inverse matrix of $\Xi$ is also block diagonal $\Xi^{-1} =
\mathrm{diag}(\Xi_\mathrm{sym}^{-1}$, $\Xi_\mathrm{asym}^{-1})$. Note
that the random vector $\hat{\bmath{P}}_\mathrm{asym}$ is not zero --
only its mean is zero. The matrix $\Xi_\mathrm{asym}$ is not a zero
matrix either. Finally, we can reduce equation~(\ref{eq:xi_prime_n2})
to $N(N+1)/2$-dimensional subspace using the inverse matrix and
$\bmath{P}_\mathrm{asym}=0$:
\begin{equation}
  \frac{\partial \bmath{P}'^\dagger}{\partial \theta_i} \Xi'^{-1} 
  \frac{\partial \bmath{P}'}{\partial \theta_j}
  = 
  \frac{\partial \bmath{P}_{\textrm{sym}}^\dagger}{\partial \theta_i} 
  \Xi_\mathrm{sym}^{-1} 
  \frac{\partial \bmath{P}_{\textrm{sym}}}{\partial \theta_j}.
\end{equation}
We complete the proof of the theorem summarised as follows. 

\subsubsection{Theorem}
The Fisher matrix of $N$ Gaussian random fields with classical
approximation (equation~\ref{eq:fisher_matrix_tr}) is equal to the
following Fisher matrix, when the matrix $\tilde{\Sigma}$ is a real
symmetric matrix (which is always possible by transforming the field
variables if necessary):
 \begin{equation}
 \label{eq:fisher_covp_sym}
  F_{ij} = \frac{1}{2} \int \frac{d^3 x d^3 k}{(2\pi)^3}
    \frac{\partial \bmath{P}}{\partial \theta_i}
    \mathrm{Cov}(\hat{\bmath{P}}_\mathrm{sym}, \hat{\bmath{P}}_\mathrm{sym})^{-1}
    \frac{\partial \bmath{P}}{\partial \theta_j},
\end{equation}
where $\bmath{P}(\bmath{k})$ is a $N(N+1)/2$-dimensional vector of
auto- and cross-power spectra $P_{ab} (a \le b)$, and the covariance
matrix can be calculated from a vector of random variables
$\hat{\bmath{P}}_\mathrm{sym}$,\footnote{Subtraction of the noise term
  is not necessary but the power spectra estimator is often written
  this way; Covariance matrix does not change by subtracting constant
  values.}
\begin{equation}
  \label{eq:Pab_sym}
  \hat{P}_{ab}^\mathrm{sym} = 
    \frac{1}{2}(\delta_a \delta_b^* + \delta_b \delta_a^*) - N_{ab}
\end{equation}
defined by $N$ random Gaussian variables $\delta_a(\bmath{k},
\bmath{x})$
(equations~\ref{eq:cov_delta}--\ref{eq:cov_delta_star}). The covariance
matrix can be written in terms of
  $\langle \delta_a \delta_b^* \rangle 
     = \tilde{\Sigma}_{ab}(\bmath{k}, \bmath{x}) 
     = P_{ab}(\bmath{k}) + N_{ab}(\bmath{x})$,
using the Wick theorem. Note that $\delta_a\delta_b^*$ and
$\delta_b\delta_a^*$ have the same expectation value, $\langle
\delta_a \delta_b^* \rangle = \langle \delta_b \delta_a^* \rangle =
P_{ab}$, but are different random variables; symmetrization in
equation~(\ref{eq:Pab_sym}) is necessary.

\subsubsection{An example of power spectra covariance matrix}
\label{sec:exaple_covp}
This is an example of the covariance matrix
$\mathrm{Cov}(\hat{\bmath{P}}, \hat{\bmath{P}})$ for two fields:
\begin{equation}
  \mathrm{Cov}(\hat{P}_{11}, \hat{P}_\mathrm{11}) = 
    \tilde{\Sigma}_{11}^2
\end{equation}
\begin{equation}
  \mathrm{Cov}(\hat{P}_{11}, \hat{P}_\mathrm{12}) = 
    \tilde{\Sigma}_{11}\tilde{\Sigma}_{12}
\end{equation}
\begin{equation}
  \mathrm{Cov}(\hat{P}_{11}, \hat{P}_\mathrm{22}) = 
    \tilde{\Sigma}_{12}^2
\end{equation}
\begin{equation}
  \mathrm{Cov}(\hat{P}_{12}, \hat{P}_\mathrm{12}) = 
    \left( \tilde{\Sigma}_{11}\tilde{\Sigma}_{22} + \tilde{\Sigma}_{12}^2 \right)/2
\end{equation}
\begin{equation}
  \mathrm{Cov}(\hat{P}_{12}, \hat{P}_\mathrm{22}) = 
    \tilde{\Sigma}_{12}\tilde{\Sigma}_{22}
\end{equation}
\begin{equation}
  \mathrm{Cov}(\hat{P}_{22}, \hat{P}_\mathrm{22}) = 
  \tilde{\Sigma}_{22}^2
\end{equation}

Similar examples are in \citetalias{2004MNRAS.347..255B} and
\citet{2009MNRAS.397.1348W}. They have factor of 2 larger covariance
matrix instead of a factor of $1/2$ in
equation~(\ref{eq:fisher_covp_sym}), which results in the same Fisher
matrix. The location of the factor of 2 only reflects the definition
of `one Fourier mode', which doesn't change the overall equation.

For galaxy density contrast
$\delta_g^s$ and transformed line-of-sight velocity
$\tilde{u}^s=\mathrm{i}u^s$ (equation~\ref{eq:u_tilde}), which give
real-number power spectra, the $\tilde{\Sigma}$ matrix elements are:
\begin{equation}
  \tilde{\Sigma}_{11} = P^s_{gg}(\bmath{k}) + n_g^{-1}(\bmath{x}),
\end{equation}
\begin{equation}
  \tilde{\Sigma}_{12} = P^s_{gu}(\bmath{k}),
\end{equation}
\begin{equation}
  \tilde{\Sigma}_{22} = P^s_{uu}(\bmath{k}) + 
                       n_u^{-1}(\bmath{x}) \sigma_\mathrm{u-noise}(\bmath{x}).
\end{equation}

The Same equation holds for galaxy density $\delta_g$ and velocity
gradient $u'$, using the conversion of power spectra in
Section~\ref{sec:examples_isomorphism}. As a corollary of
Appendices~\ref{sec:fisher_matrix_transformation}--\ref{sec:fisher_matrix_covp},
we show that our Fisher matrix formula
equation~(\ref{eq:fisher_matrix}) is exactly the same Fisher matrix
formula used in \citetalias{2004MNRAS.347..255B}, which is written
with a covariance matrix of density and velocity gradient power
spectra.

\bsp
\label{lastpage}
\end{document}